\documentclass{aa}

\usepackage{bm}

\newcommand{\dif}{\mathrm{d}}

\newcommand{\R}{\mathbb{R}}
\newcommand{\norm}[1]{\left\|#1\right\|}
\newcommand{\abs}[1]{\left|#1\right|}	

\DeclareMathOperator*{\argmin}{argmin}

\newcommand{\Set}[2]{\left\{\,#1:\, #2\,\right\}}

\newcommand{\Covf}{{\bm \Sigma^{\bm f}}}
\newcommand{\Noicov}{{\bm \Sigma^{\bm y}}}

\newcommand{\f}{\bm{f}}	
\newcommand{\fmapvec}{ \bm{f}^{\text{MAP}} }	
\newcommand{\fmap}{ f^\text{MAP} }

\newcommand{\y}{\bm{y}}
\newcommand{\g}{\bm{g}}
\newcommand{\G}{\bm{G}}
\newcommand{\U}{\bm{U}}
\newcommand{\V}{\bm{V}}
\newcommand{\I}{\bm{I}}
\newcommand{\Z}{\bm{Z}}
\newcommand{\vmu}{ {\bm{\mu}} }		

\newcommand{\0}{\bm{0}}	
\newcommand{\losvd}{\mathcal{L}}	

\newcommand{\region}{C^\alpha} 
\newcommand{\range}[1]{\lbrace 1, \ldots, #1 \rbrace}	

\newcommand{\flow}{ \bm{f}^\mathrm{low} }
\newcommand{\fupp}{ \bm{f}^\mathrm{upp} }

\newcommand{\Fcilow}{\bm L^\mathrm{low}}	
\newcommand{\Fciupp}{\bm L^\mathrm{upp}}	

\newcommand{\fcilow}{L^\mathrm{low}}
\newcommand{\fciupp}{L^\mathrm{upp}}

\newcommand{\x}{ {\bm{x}} }		
\newcommand{\obs}{ {\bm{y}} }	

\newcommand{\blanket}{\bar{\bm B}}	
\newcommand{\ssrmap}{\bm L^\text{MAP}}	

\newcommand{\normlap}{\Delta^\text{norm}} 
\newcommand{\disnormlap}{\bm \Delta^\text{norm}} 

\usepackage{graphicx}
\usepackage{txfonts}
\graphicspath{{images/}}

\usepackage{xcolor}
\usepackage{comment}
\usepackage[normalem]{ulem}

\ifdefined\markchanges
	\newcommand{\added}[1]{{\color{teal}#1}}
	\newcommand{\removed}[1]{{\color{red}#1}}
\else
	\newcommand{\added}[1]{#1}
	\newcommand{\removed}[1]{}
\fi

\begin{document}

\defcitealias{BoeAlfNeuMarLea20e}{B20}

	\title{Uncertainty-Aware Blob Detection with an Application to Integrated-Light Stellar Population Recoveries}

	\author{Fabian Parzer\inst{*,1}
    \and Prashin Jethwa\inst{*,2}
		\and Alina Boecker\inst{3,4}
		\and Mayte Alfaro-Cuello\inst{5,6}
		\and Otmar Scherzer\inst{1,7,8}
		\and \\Glenn van de Ven\inst{2}
	}

	\institute{
	\inst{*} \textit{Both authors contributed equally to this work.}\\
	\inst{1} Faculty of Mathematics, University of Vienna, Oskar-Morgenstern-Platz 1, A-1090 Vienna \\
	\inst{2} Department of Astrophysics, University of Vienna, T\"urkenschanzstra{\ss}e 17, A-1180 Vienna, Austria \\
	\inst{3} Max-Planck Institut f\"{u}r Astronomie, K\"{o}nigstuhl 17, D-69117, Heidelberg, Germany \\
	\inst{4} Instituto de Astrof\'{i}sica de Canarias, C/ V\'{i}a L\'{a}ctea s/n, E-38205 La Laguna, Spain \\
	\inst{5} Facultad de Ingenier\'{i}a y Arquitectura, Universidad Central de Chile, Av. Francisco de Aguirre 0405, La Serena, Coquimbo, Chile \\
	\inst{6} Space Telescope Science Institute, 3700 San Martin Drive, Baltimore, MD 21218, USA \\
	\inst{7} Johann Radon Institute for Computational and Applied Mathematics (RICAM), A-4040 Linz \\
	\inst{8} Christian Doppler Laboratory for Mathematical Modeling and Simulation of Next Generations of Ultrasound Devices (MaMSi), Oskar-Morgenstern-Platz 1, A-1090 Vienna \\
	\email{prashin.jethwa@univie.ac.at}
	}

   \date{\today}


  \abstract
   {Blob detection is a common problem in astronomy. One example is in stellar population modelling, where the distribution of stellar ages and metallicities in a galaxy is inferred from observations. In this context, blobs may correspond to stars born in-situ versus those accreted from satellites, and the task of blob detection is to disentangle these components. A difficulty arises when the distributions come with significant uncertainties, as is the case for stellar population recoveries inferred from modelling spectra of unresolved stellar systems. There is currently no satisfactory method for blob detection with uncertainties.}
   {We introduce a method for uncertainty-aware blob detection developed in the context of stellar population modelling of integrated-light spectra of stellar systems.}
   {We develop theory and computational tools for an uncertainty-aware version of the classic Laplacian-of-Gaussians method for blob detection, which we call ULoG. This identifies significant blobs considering a variety of scales. As a prerequisite to apply ULoG to stellar population modelling, we introduce a method for efficient computation of uncertainties for spectral modelling. This method is based on the truncated Singular Value Decomposition and Markov Chain Monte Carlo sampling (SVD-MCMC).
	 }
   {We apply the methods to data of the star cluster M54. We show that the SVD-MCMC inferences match those from standard MCMC, but are a factor 5-10 faster to compute. We apply ULoG to the inferred M54 age/metallicity distributions, identifying between 2 or 3 significant, distinct populations amongst its stars.}
   {}


	 \titlerunning{Uncertainty-Aware Blob Detection for Stellar Populations}
	 \authorrunning{F. Parzer et al.}
   \maketitle
%

\section{Introduction}\label{sec:introduction}

Identifying distinct structures in a distribution is a common problem in astronomy. When the distribution under consideration is an observed image, this problem is called source detection \citep[see][for a review]{Masias12}. In many cases, however, the distribution is not an observed image, but instead is itself the result of some earlier inference problem. Consider the following examples:
\begin{enumerate}
  \item \removed{Graviational}\added{Gravitational} lensing, where a mass distribution is inferred from an image which displays lensing effects and structures such as dark matter halos are identified in the mass distribution \citep[e.g.][]{Despali22}.
  \item Orbit-superposition modelling of galaxies, where an orbit distribution is inferred from stellar kinematic data and structures, i.e. dynamical components of the galaxy, are identified in the orbit distribution \citep[e.g.][]{Zhu22}.
  \item Stellar population modelling of unresolved stellar systems, where a distribution over stellar ages and metallicities is inferred from spectral modelling and structures, i.e. distinct populations, are identified there within \citep[e.g.][]{BoeAlfNeuMarLea20e}.
\end{enumerate}
A theme connecting all three examples - and the important distinction with respect to observed images - is that the distribution may come with significant uncertainties as a result of the initial inference.

Let's consider stellar population modelling in more detail. Non-parametric modelling of integrated-light spectra is a flexible way to recover stellar population distributions. In this approach, libraries of several hundred template spectra of simple stellar populations (SSPs) are superposed to model observed spectra \citep[e.g.][]{CapEms04,CidFernandes05,Ocvirk06,Tojeiro07,Koleva08}. This approach offers the flexibility to reconstruct complex star formation histories (SFHs) where metallicity and alpha abundances \citep{Vazdekis15} can vary freely with stellar age. In cases where stellar systems host several distinct populations, nonparametric models can recover this complexity \citep{BoeAlfNeuMarLea20e}. A difficulty with nonparametric models is that the high-dimensionality has hindered robust uncertainty quantification. An alternative is to use parametric models, where the SFH is characterised with at most a few dozen parameters \citep[e.g.][]{Gladders13,Simha14,Carnall18}. Whilst uncertainty quantification is easier here, parametric models impose strong assumptions which, if invalid, may bias inferences \citep{Lower20} and lead to underestimated uncertainties \citep{Leja19}.

In this work we describe a method for uncertainty quantification for nonparametric stellar population modelling. Furthermore, we introduce a general method for \emph{blob detection} which can identify distinct blobs in reconstructed distributions with uncertainties. In the context of stellar populations, \removed{these}\added{distinct} blobs will correspond to groups of stars with different origins -- e.g. those formed in-situ within a galaxy versus those accreted from merged, destroyed satellites \citep{white_rees_78}. Due to the extended star-formation histories of galaxies and their satellites, blobs may partially overlap in age/metallicity space, therefore blob detection in this context is distinct from the task of image segmentation.

Blob detection has traditionally been studied in the case where the underlying image is known. When the image of interest has to be reconstructed from noisy measurements, the result of the blob detection itself becomes uncertain. Currently, the dominating approach for uncertainty quantification in imaging is Bayesian statistics \citep{Han93,KaiSom07,DasStu17}. While current research mostly focuses on the development of computationally efficient methods for generic uncertainty quantification, in particular MCMC \citep{Bar12,MarWilBurGha12,ParDavSee21}, the specific task of combining uncertainty quantification with blob detection has not been explicitly studied. \cite{RepPerWia18} proposed a method for significance tests of image structures called "Bayesian uncertainty quantification by convex optimisation" (BUQO). \cite{PriCaiMcEPerKit20} proposed a similar test that detects a structure in an uncertain image by checking whether a suitable surrogate image is contained in the posterior credible region. See also the related works by Cai et al. (\citeyear{CaiPerMcE18a} \citeyear{CaiPerMcE18}) on uncertainty quantification for radio interferometric imaging, and by Price et al. (\citeyear{PriCaiMcEPerKit20}, \citeyear{PriMcECaiKitWal21}) on sparse Bayesian mass mapping. The cited methods are computationally very efficient, but do not perform automatic scale-selection, assuming instead that the structure of interest is well-localised in a user-specified image region. This makes them also less suitable for the case where structures might overlap, as is the case with stellar populations.

We have based our approach for blob detection on the classic Laplacian-of-Gaussians method \citep{Lin98}. This method is able to determine the location and scale of blob-like structures in an arbitrary image. Although the underlying ideas of the Laplacian-of-Gaussians have been modified and extended in subsequent work \citep[see e.g. the reviews by][]{MikTuySchZisMat05,Lin13,LiWanTiaDin15}, we have based our approach on the original formulation due to its conceptual simplicity and straightforward mathematical definition.

\removed{This paper is structured as follows. Section \ref{sec:background} introduces the problem of stellar population modelling of integrated-light spectra. Section \ref{sec:lapgau} outlines the existing Laplacian-of-Gaussians \removed{menthod}\added{method} for blob detection, while Section \ref{sec:uncertainty_aware_blob_detection} describes our extension of this method to account for uncertainties. As a prerequisite to implement this method, Section \ref{sec:svd_mcmc} describes a computationally efficient MCMC method to quantify uncertainties for this problem. In Section \ref{sec:m54} we apply the method to MUSE data of the star cluster M54, before concluding in Section \ref{sec:conclusions}. Ac\added{c}ompanying code used for all the analysis is available online\footnote{\url{https://github.com/prashjet/uq4pk/}}.}

\begin{figure}
   \centering
   \includegraphics[width=\columnwidth]{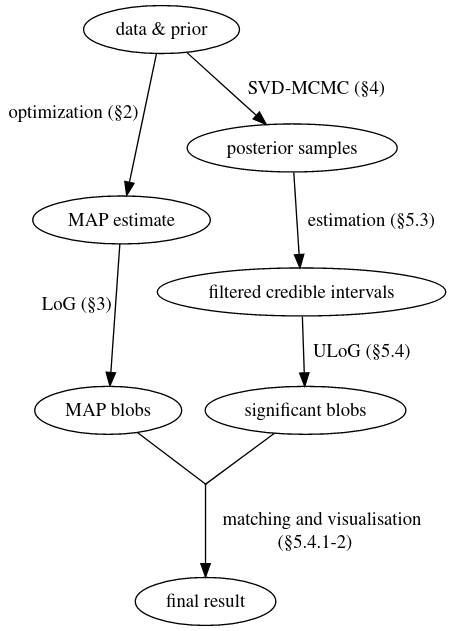}
      \caption{\added{High-level overview of our method for uncertainty-aware blob detection.}}
   \label{fig:overview}
\end{figure}

\added{
\subsection{Overview}\label{ssec:overview}

An overview of our technique is shown in figure \ref{fig:overview}. Given an observed spectrum and a choice of prior, a set of reference blobs is determined from the maximum-a-posteriori (MAP) estimate. Using posterior sampling we then determine those components of the image that are distinguishable with confidence. These \emph{significant blobs} are then combined with the reference blobs using a matching procedure, yielding a final visualisation that shows which blobs are significant, whether blobs are separable with confidence, and which blobs are explainable by noise alone. An example of this is shown in figure \ref{fig:significant_blobs}.
}

\begin{figure}
   \centering
   \includegraphics[width=\hsize]{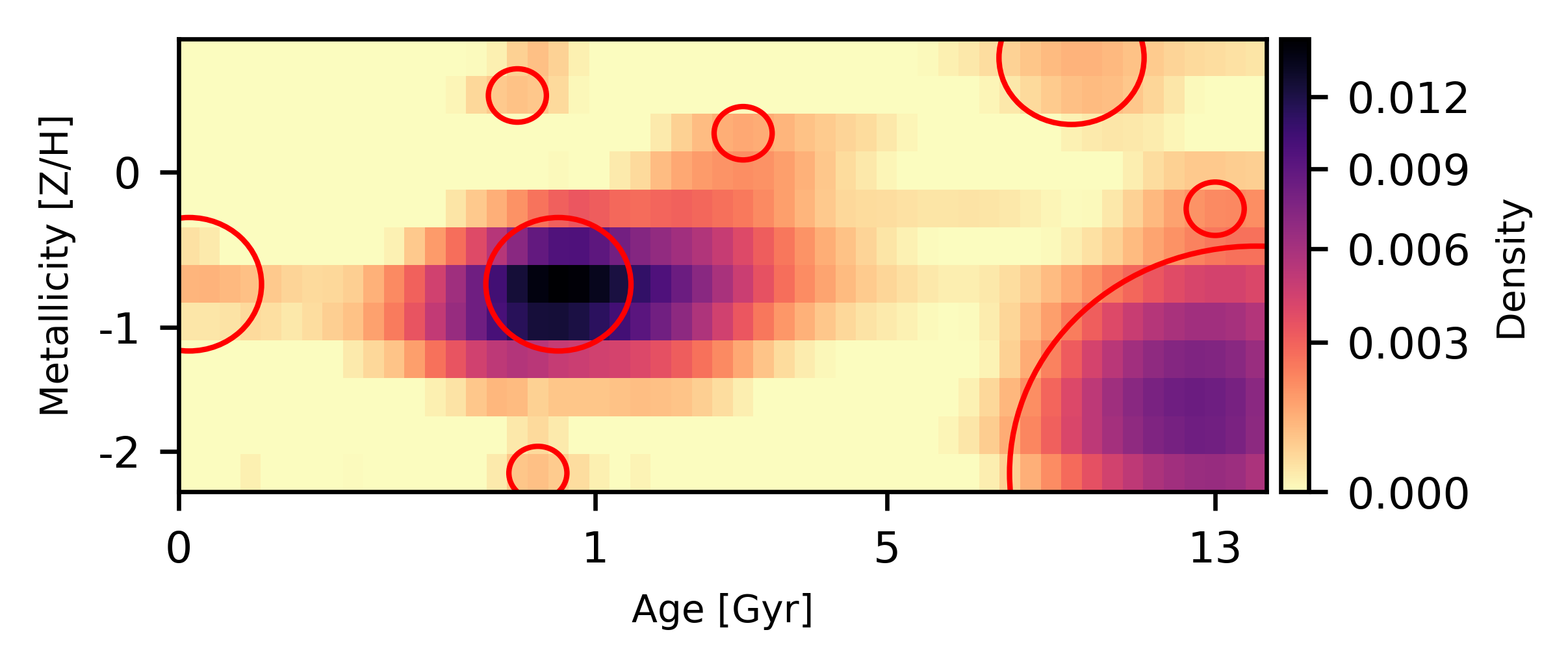}
      \caption{\added{Results of the LoG method (see section \ref{sec:lapgau}) applied to the MAP estimate of the stellar distribution function. The detected blobs are indicated by red circles.}}
         \label{fig:log_demo}
   \end{figure}

\begin{figure}
   \centering
   \includegraphics[width=\hsize]{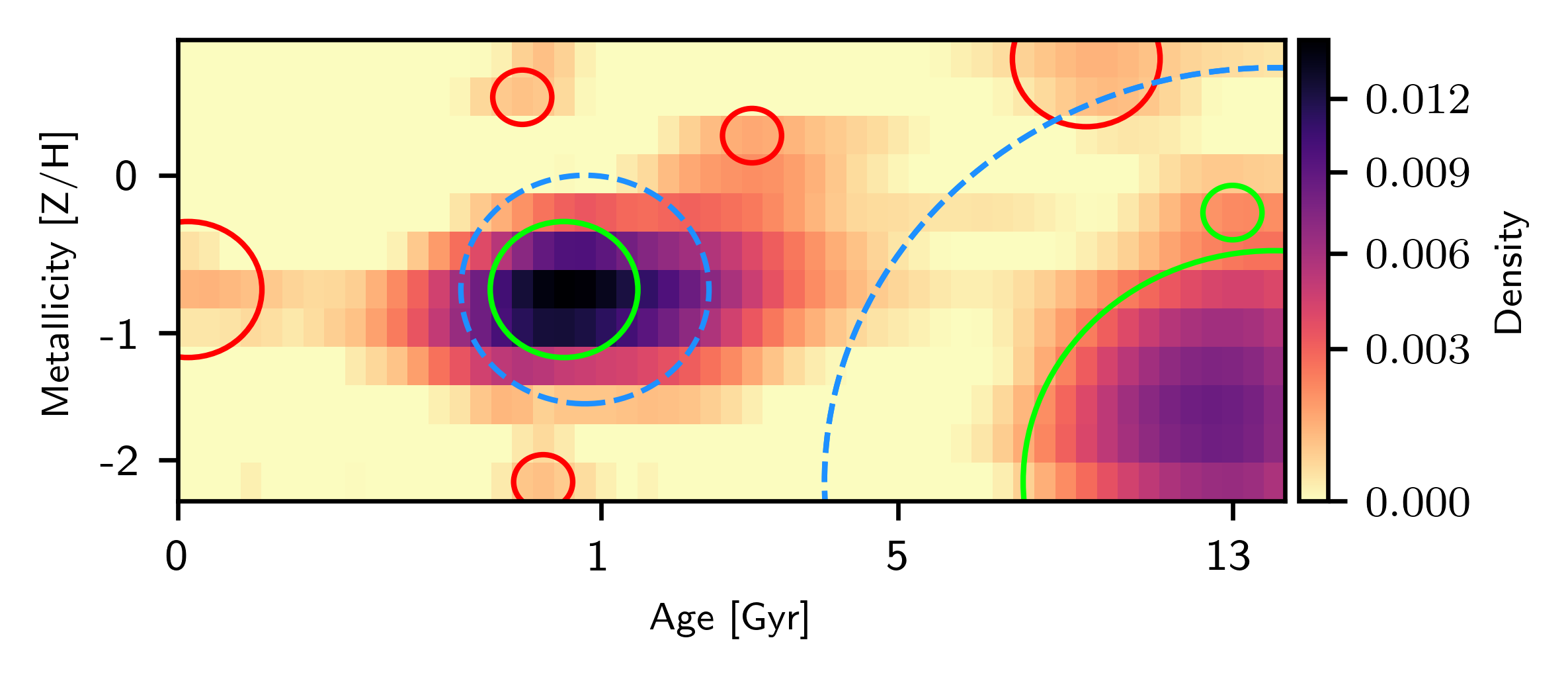}
      \caption{\added{Results of the proposed ULoG method to detect blobs with 95\% credibility. The mock data and MAP estimate are the same as in figure \ref{fig:log_demo}. The significant blobs are visualised by dashed blue circles, their size corresponding to localisation uncertainty. Blobs in the MAP estimate that correspond to significant structures are marked in green, blobs that might be explainable by noise are marked in red (see section \ref{sssec:visualise}).}}
         \label{fig:significant_blobs}
   \end{figure}

\added{In more detail: First, we obtain the MAP estimate through optimisation (see equation \eqref{eq:map} and the surrounding discussion). We then apply the Laplacian-of-Gaussians (LoG) method (described in section \ref{sec:lapgau}) to determine the position and scale of blobs in the MAP image. An example of this is visualised in figure \ref{fig:log_demo}.}

\added{When accounting for uncertainties, however, not all of these blobs are significant. To quantify the uncertainty in our estimate, we generate posterior samples using MCMC. Standard MCMC algorithms require several hours for our problem, so we developed a new technique for efficient sampling based on dimensionality reduction, called SVD-MCMC (described in section \ref{sec:svd_mcmc}).}

\added{To assess the significance of the image blobs from the posterior samples we use a tube-based approach: First we compute so-called filtered credible intervals (see sections \ref{subsec:filtered_credible_intervals} and \ref{subsec:fcis_from_samples}) that, for a given scale, consist of an upper and lower bound in which the true image lies with a high probability. We can then determine significant structures by "spanning a blanket", i.e. computing the image with the least amou\added{n}t of blobs that still lies within the credible intervals (see section \ref{subsec:ulog}). Since the uncertainty of fine image features is usually different than the uncertainty of large-scale structures, this has to be repeated for multiple scales. Applying the LoG method to the resulting stack of blankets then allows us to determine which image blobs are significant. Hence, we call this technique "Uncertainty-aware Laplacian-of-Gaussians" (ULoG).
}

\added{To visualise the results of ULoG as in figure \ref{fig:significant_blobs}, we use the MAP image as a reference and highlight which blobs detected by the LoG method - visualised in figure \ref{fig:log_demo} - are significant and which are not. The details of this procedure are described in sections \ref{sssec:matching} and \ref{sssec:visualise}.}

\section{Problem setup}\label{sec:background}


Our goal is to robustly identify distinct blobs in stellar population distribution functions by modelling integrated-light spectra. The distribution function $f$ is a function of stellar \removed{age $t$ and metallicity $z$}\added{metallicity $z$ and age $t$}, and it is physically constrained to be non-negative, i.e.
\begin{equation*}
f(\removed{t, }z\added{, t}) \geq 0.
\end{equation*}
This can be related to an observed spectrum via simple stellar population (SSP) templates $S(\lambda,\removed{t,}z\added{,t})$ which encode the spectra of single-age and single-metallicity stellar populations. The spectrum of a composite stellar population is a combination of SSPs weighted by $f$. Assuming that all stars in this composite population share a line-of-sight velocity distribution (LOSVD) $\losvd(v)$, then the observed spectrum is given by the composite rest-wavelength spectrum convolved by $\losvd(v)$. Putting this all together, the observation operator $\mathcal{G}$ is given by
\begin{equation}
\mathcal{G}(f, \losvd) =  \int
  \frac{\losvd(v)}{1+v/c}
  \left[
  \int
  \int
  S\left(\frac{\lambda}{1+v/c},\removed{t,}z\added{,t}\right)
  f(\removed{t, }z\added{, t}) \; \mathrm{d}t \; \mathrm{d}z
  \right]
  \; \mathrm{d}v.
\label{eq:obs_operator}
\end{equation}
The observed data $y$ is given by the addition of noise $\epsilon$ which is wavelength-dependent,
\begin{equation}
y(\lambda) =  \mathcal{G}(f, \losvd) + \epsilon(\lambda).
\label{eq:ybar}
\end{equation}
Since the goal of this work is to develop methods to robustly identify structures in $f$, we will fix the LOSVD $\losvd$ in our recoveries.
Whilst equations \eqref{eq:obs_operator} and \eqref{eq:ybar} relate continuous functions $y, f$ and $\losvd$, in reality observations are made at a finite number of wavelengths $\lambda$. We therefore instead work with the discretised version of equation \eqref{eq:ybar}, for fixed $\losvd$, i.e.
\begin{equation}
\y = \G \f + \bm{\epsilon}.
\label{eq:ybar_discrete}
\end{equation}
Details of converting between continuous and discretised equations are in section \ref{subsec:discretisation}. The noise is typically assumed to be distributed as a multivariate normal with some covariance $\Noicov$,
\begin{equation*}
\bm{\epsilon} \sim \mathcal{N}(0, \Noicov),
\end{equation*}
where we assume that an estimate of $\Noicov$ is available as is typically output from data-reduction pipelines. This completes the definition of our forward model. For the time being, we ignore several other physical ingredients which go into complete spectral models -- e.g. LOSVD modelling, gas emission, reddening and calibration related continuum distortions. We discuss these effects further in Section \ref{subsec:further_applications}.

Non-parametric recovery of $f$ from noisy observations $y$ is an ill-conditioned inverse problem. This means that noise in the observations is amplified in the recovery, leading to unrealistically spiky solutions. Solutions are found by minimising the mismatch between data and model as quantified by $\chi^2$,
\begin{equation*}
\chi^2 = \norm{\y - \G \f}_\Noicov^2.
\end{equation*}
The \emph{maximum likelihood} (ML) solution is the one which minimises the $\chi^2$ mismatch, i.e.
\begin{equation*}
\f_\mathrm{ML} = \argmin_{\f} \chi^2.
\end{equation*}
To counter ill-conditioning, regularisation can be used to promote smoothness in the solution. In practice this means that rather than finding the ML solution,
one instead finds the \emph{maximum a posteriori} (MAP) solution by adding a penalty term to the objective function,
\begin{equation}
\f_\mathrm{MAP} = \argmin_{\f} \left( \chi^2 + \beta \mathcal{R} \right).
\label{eq:map}
\end{equation}

The penalty consists of a functional $\mathcal{R}(\f)$ which quantifies the smoothness of $\f$ multiplied by a tunable scalar $\beta$ called the regularisation parameter. The functional is non-negative, and decreases for increasingly smooth $\f$.  When $\beta=0$, the MAP and the ML solutions coincide, while in the limit of very large $\beta$, $\f_\mathrm{MAP}$ will be very smooth at the expense of no longer well reproducing the data. By appropriately tuning $\beta$ to some intermediate value however the MAP solution will exhibit a good data-fit and a desired degree of smoothness. MAP recoveries are commonly used in stellar population modelling: \citet{Ocvirk06} give a detailed description of the theory behind regularisation in this context, and it is available in the popular spectral modelling software Penalised Pixel-Fitting \citep[pPXF,][]{Cappellari17}.

We will not address the question of how to tune the regularisation parameter $\beta$ in this work. Several other suggestions have been made elsewhere. For example, the pPXF documentation \footnote{\url{https://pypi.org/project/ppxf/}} offers the guideline to first perform an un-regularised fit, then increase $\beta$ till the data mismatch increases to $\Delta \chi^2 = \sqrt{2n}$ where $n$ is the size of the spectrum. Further well-known parameter choice rules are the L-curve method \citep[][]{HanOle93} and Morozov's discrepancy principle \citep[][provides a theoretical discussion]{Gro83}. Alternative Bayesian approaches for choosing the regularisation parameter are discussed in section \ref{subsec:svd_mcmc_discussion}. While the choice of regularisation parameter can affect the number of blobs visible in a recovered $\f_\mathrm{MAP}$, in this work we restrict ourselves to the question of quantifying the uncertainty in the solution at fixed $\beta$ and then using these uncertainty estimates to identify distinct blobs.

Regularisation can also be understood from a Bayesian perspective.
Under this interpretation, equation \eqref{eq:map} comes from the combination of a likelihood function
\begin{equation*}
p(\y | \f) \propto e^{-\frac{1}{2} \chi^2}
\end{equation*}
with a prior probability density function
\begin{equation}
p(\f) \propto e^{-\frac{1}{2} \beta \mathcal{R}}.
\label{eq:prior}
\end{equation}
Bayes' theorem tells us that the posterior probability is proportional to the product of the likelihood and prior, i.e.
\begin{equation*}
p(\f|\y) \propto e^{-\frac{1}{2} \left(
  \chi^2 + \beta \mathcal{R}
  \right)},
\end{equation*}
from which we can see that equation \eqref{eq:map} picks out the solution $\f_\mathrm{MAP}$ which maximises the (log) posterior probability function. This is the origin of the name \emph{maximum a posteriori}.

In this work, we assume a Gaussian prior on $\f$ that is truncated to the nonnegative orthant, since the density $\f$ cannot take negative values. That is, we make the choice
\begin{equation}
p(\f) \propto \begin{cases}
e^{- \frac{\beta}{2} \norm{\f}_\Covf^2}, & \text{if } \f \geq 0, \\
0, & \text{otherwise}.
\end{cases}\label{eq:truncated_gaussian_prior}
\end{equation}
In the spirit of regularisation theory, we use the prior covariance $\Covf$ to encode smoothness assumptions on $\f$ \citep{CalSom18}. Details on our choice for $\Covf$ are given in section \ref{subsec:discretisation}.

Viewed from this Bayesian perspective, we can explicitly see that regularisation incorporates prior information into the solution, and the regularisation parameter $\beta$ controls the strength of the prior. Whilst a MAP solution may be considered Bayesian in the sense that it assumes a prior, it is nevertheless a single point estimate which on its own provides no information about uncertainty of the solution. In this work, we will introduce an efficient method to estimate this uncertainty based on dimensionality-reduced MCMC sampling.

\begin{figure*}
   \centering
   \includegraphics[width=\textwidth]{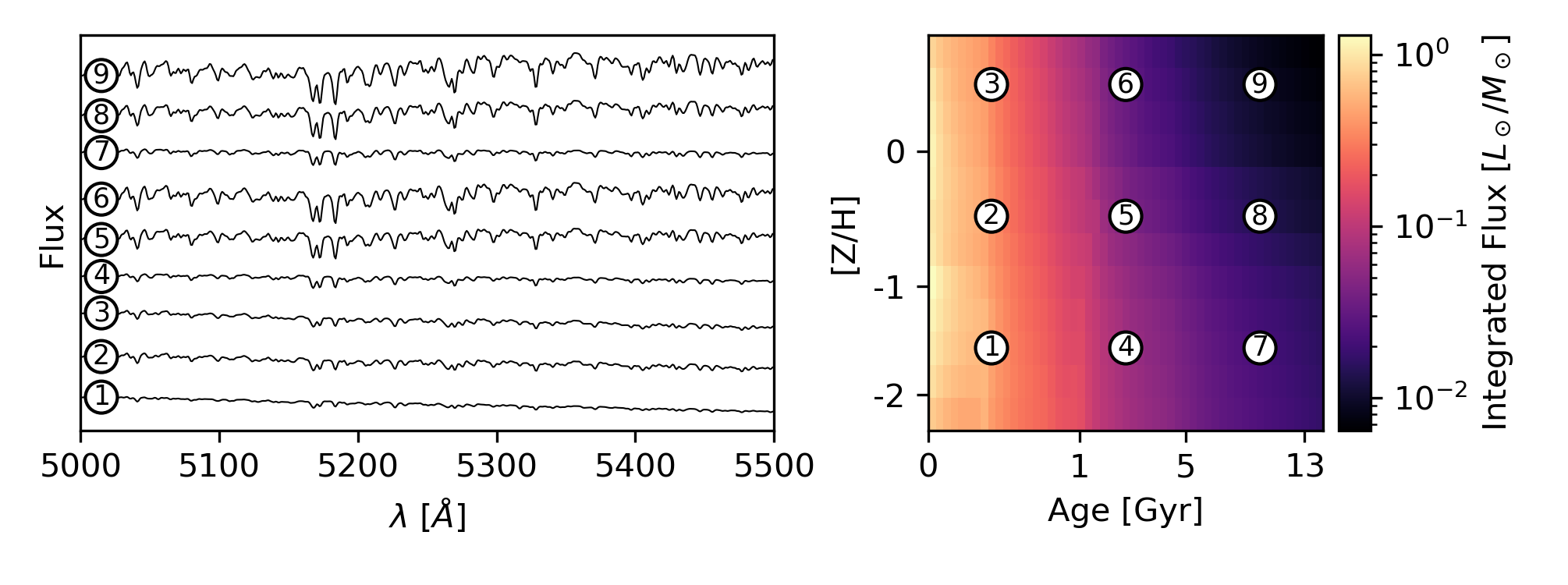}
      \caption{Examples of SSP templates.
			Left panel shows 9 representative models from the MILES SSP library. These are labelled 1-9, corresponding to the position on the age-metallicity plane as shown in the right panel. The colorbar shows the integrated flux of the full grid of all 636 SSP templates as a function of age and metallicity.
Younger stars are hotter and therefore brighter, explaining why integrated flux is a strong function of SSP age. Integrated flux depends more weakly on metallicity. For fixed age, we see that SSPs with greater metallicity have stronger absorption features e.g. compare spectra 7, 8 and 9 in the left panel.}\label{fig:ssp_templates}
\end{figure*}

The use of regularisation complicates another question commonly faced in stellar population modelling: the choice of whether to reconstruct \emph{mass-weighted} or \emph{light-weighted} quantities.
These are two options for the parametrisation of $\f$ connected to the normalisation of SSP templates $S$ in equation \eqref{eq:obs_operator}. If the SSPs are normalised to represent the flux of a unit \emph{mass} of stars, $\f$ is said to be mass-weighted. In this case, the integrated flux of a stellar population is given by
\begin{equation*}
\text{Flux}(\removed{t,}z\added{,t}) = \int S(\lambda,\removed{t,}z\added{,t}) \; \mathrm{d}\lambda.
\end{equation*}
SSP integrated flux varies strongly with age and, to a lesser extent, metallicity (Figure \ref{fig:ssp_templates}). As an alternative to mass-weighting, the SSP templates $S$ in equation \eqref{eq:obs_operator} may be normalised to all have the same integrated flux, in which case the recovery is said to be light-weighted. Maximum-likelihood mass-weighted and light-weighted solutions should be self-consistent in the sense that
\begin{equation*}
\textnormal{light-weighted} \; \f_\mathrm{ML}(\removed{t,}z\added{,t}) = \text{Flux}(\removed{t,}z\added{,t}) \times \textnormal{mass-weighted} \; \f_\mathrm{ML}(\removed{t,}z\added{,t}).
\end{equation*}
This consistency does not hold for MAP solutions. That is to say, the decision to regularise mass-weighted or light-weighted $f$ will affect the result.

Is there a preferred option between regularised mass- or light-weighted recoveries? We think there is no single objective correct answer to this question. Physically, one may imagine a mass of gas converting to stars at a rate varying smoothly with time, whence regularised mass-weights might seem to be the preferred option. Given that integrated SSP flux varies smoothly with age and metallicity however (see Figure \ref{fig:ssp_templates}), mass-weighted smoothness would also imply light-weighted smoothness. Whilst both options may be reasonably justified under physical considerations, we raise this discussion here since \emph{numerically} the situation is more complicated. For the MCMC based uncertainty quantification method we will present in Section \ref{sec:svd_mcmc}, we find that sampling a model with regularised mass-weights is numerically unstable; see that section for further details. In order to facilitate comparison between all methods presented in this work, we will therefore only consider regularised light-weighted recoveries throughout. We also remark that this choice can be interpreted as a "linear functional strategy" \citep[see][]{And86} for reducing the ill-posedness of the reconstruction.

\subsection{Discretisation}\label{subsec:discretisation}

Here we outline the details of converting from continuous variables $f$, $y$, $S$ etc. in equation \eqref{eq:ybar}, to the discretised version for a fixed LOSVD, i.e. equation \eqref{eq:ybar_discrete}. We assume that data is available at $n$ wavelengths
\begin{equation*}
\y = [y(\lambda_1), ..., y(\lambda_n)],
\end{equation*}
where typical values of $n$ are around 1000. The domains of $t$ and $z$ can be discretised into $T$ and $Z$ intervals respectively. In pratice, the age and metallicity discretisations are set by the choice of SSP library. The MILES library which we use has $\removed{(T,Z)=(53,12)}\added{(Z,T)=(12, 53)}$.
We next introduce vectors $\g_{ij}$ equal to the SSP template of age $t_i$ and metallicity $z_i$, pre-convolved by the (known) LOSVD, and multiplied by the discretisation interval widths, i.e.
\begin{align*}
&\g_{ij} = [g_{ij}(\lambda_1), ..., g_{ij}(\lambda_n)], \\
&g_{ij}(\lambda) =
  \Delta t_i \Delta z_j
  \int
  \frac{L(v)}{1+v/c} S\left(\frac{\lambda}{1+v/c},\removed{t_i,}z_j\added{, t_j}\right) \mathrm{d}v.
\end{align*}
Equation \eqref{eq:ybar} now reduces to
\begin{equation}
\y = \sum_{i=1}^{Z} \sum_{j=1}^{T} \g_{ij} f_{ij} + \bm{\epsilon},
\label{eq:y_F}
\end{equation}
i.e. the data is simply a sum of (pre-convolved) templates $\g_{ij}$ weighted by the 2D discretised distribution function $f_{ij}$, plus noise $\bm{\epsilon}$.

We correspondingly collect the templates $\g_{ij}$ into a 3D tensor $\G$ of shape $(n, \removed{T, }Z\added{, T})$. Equation \eqref{eq:y_F} can then be rewritten using tensor notation, i.e.
\begin{displaymath}
\y = \G \f + \bm{\epsilon},
\end{displaymath}
Corresponding to our choice to represent the image $\f$ as a matrix, the prior covariance $\Covf$ in \eqref{eq:truncated_gaussian_prior} is given by a 4-tensor, where the entry $\Covf_{ij, kl}$ is equal to the covariance of the pixel intensities $f_{ij}$ and $f_{kl}$. As we remarked above, different choices for $\Covf$ correspond to different regularisation strategies.
In this paper, we use the Ornstein-Uhlenbeck covariance
\begin{equation}
\Sigma^{\f}_{ij,kl} = e^{-(\abs{i-k} + \abs{j - l}) / h}.\label{eq:ornstein_uhlenbeck}
\end{equation}
This choice corresponds to the intuitive assumption that neighbouring pixels in $\f$ are positively correlated, and that the amount of correlation decreases exponentially with distance. The parameter $h > 0$ is sometimes called the ``correlation length'', as it controls the rate of this decrease.

\subsection{Mock data}
\label{ssec:mock_data}

Throughout Sections \ref{sec:lapgau} - \ref{sec:uncertainty_aware_blob_detection}, we will use mock data to illustrate our methods. That is, we simulate a known ground truth LOSVD and distribution function $\f$, apply the observation operator and add artificial noise to generate mock, noisy data. Here we describe the specific choices taken for these steps.

For the LOSVD we will use a fourth order Gauss-Hermite expansion \citep{vdMarelFranx93} which takes parameters
\begin{equation*}
\bm \theta_v = (V, \sigma, h_3, h_4),
\end{equation*}
which correspond roughly to the mean, standard deviation, skewness and kurtosis of $L(v)$. We will generate mock-data using plausible galactic LOSVDs, namely using parameters $\bm \theta_v = [30, 100, -0.05, 0.1]$. We fix $\bm \theta_v$ to the known ground truth during recoveries.

For the SSPs, we use the MILES\footnote{\url{http://miles.iac.es/pages/ssp-models.php}} models \citep{Vazdekis10} with BaSTI isochrones for base alpha-content \citep{Pietrinferni04} and a \citet{Chabrier03} initial mass function. These models span 53 ages in the range $0.03<t/\mathrm{Gyr}<14$ and 12 values of metallicity $-2.27<z/\mathrm{[M/H]}<0.4$. Both age and metallicity values are non-uniformly spaced in these ranges. The overall 2D grid of SSP models consists of 636 template spectra. We use the wavelength range between $[4700, 6500]$ Angstroms with a sampling in velocity space of 10 km s$^{-1}$.

For the distribution functions we use random Gaussian mixtures. We generate ground-truth, light-weighted distributions $f$ as mixtures of three Gaussian blobs which are randomly sized, shaped and positioned in the (non-uniformly spaced) age and metallicity domain shown e.g. in the right panel of Figure \ref{fig:ssp_templates}. Specifically, the parameters describing the blobs are sampled in normalised co-ordinates which increase linearly from $[0,1]$ between the corners of the domain. Blob means are sampled uniformly between $[0,1]$. The per-dimension variances are sampled independently from log-uniform distributions between $[0.01,1]$ and the blob covariance matrix is constructed using these variances and a random rotation. The blob weights are sampled from log-uniform distributions in the range $[1,30]$.

Finally, for the noise model we sample independent, normally distributed noise with a constant standard deviation $\sigma^2$, i.e.
\begin{equation}
\Noicov = \mathbf{diag}\{(\sigma^2, ..., \sigma^2)\},
\label{eq:noise}
\end{equation}
where $\sigma$ is chosen so that the average signal-to-noise ratio (SNR) equals some specified value
\begin{equation*}
\frac{\norm{\bar \y}}{\sigma} = \mathrm{SNR}.
\end{equation*}
Unless otherwise specified, we will use $\mathrm{SNR} = 100$.

\section{The Laplacian-of-Gaussians method for blob detection}\label{sec:lapgau}

In this section, we recall the necessary prerequisites on the Laplacian-of-Gaussians method. It is defined in terms of the Gaussian scale-space representation, which we introduce first.

\subsection{The scale-space representation of an image}\label{subsec:scale_space}

The \emph{scale-space representation} of an image is a classic tool from computer vision, where it is used to study image structures at their intrinsic scale. For an exhaustive reference on scale-space theory in computer vision, see \citet{Lin94}.

A scale-space representation of a continuous image $f:\R^2 \to \R$ is a one-parameter family $(L_t)_{t \geq 0}$ of images with $L_0 = f$, where the continuous parameter $t \geq 0$ corresponds to physical scale. We will use the notation $L(\bm x, t) = L_t(\bm x)$ to refer to the scale-space representation as a function both in $\x$ and $t$.

One distinguishes linear and nonlinear scale space representations, where a scale space is called linear if, for all $t \geq 0$, $L_t$ is obtained from $f$ by a linear operation, and else is called nonlinear. The Gaussian scale space is arguably the most well-studied scale space, since it is the unique linear scale space that satisfies shift-invariance, scale invariance, the semigroup property and non-enhancement of local extrema (see \citet{FloTerKoeVie92}). It is defined as
\begin{displaymath}
L(\x, t) = k_t * f,
\end{displaymath}
where $k_t$ is a two-dimensional Gaussian of the form
\begin{displaymath}
k_t = \frac{1}{2 \pi t} e^{-\norm{x}^2 / (2 t)}.
\end{displaymath}

\subsection{The Laplacian-of-Gaussians method for blob detection}\label{subsec:laplacian_of_gaussians}

Given the Gaussian scale-space representation $L$ of a continuous image $f$, the \emph{Laplacian-of-Gaussians} (LoG) method \citep[see][]{Lin98} detects blob-like image structures as the local minimisers and maximisers of the scale-normalised Laplacian of $L$, which is defined as
\begin{equation}
\normlap L(\x, t) = t (\partial_{x_1}^2 + \partial_{x_2}^2) L(\x, t). \label{eq:normalized_laplacian}
\end{equation}
A local minimiser $(\x, t) \in \R^2 \times [0, \infty)$ of $\normlap L$ corresponds to a bright blob of scale $t$ with center $\x$. On the other hand, local maximisers correspond to dark blobs. For the rest of this article, we will only consider the identification of bright blobs, since as explained in section \ref{sec:introduction}, we expect them to correspond to distinct stellar populations. While alternative ways to formally define blobs in an image have been studied \citep[e.g. in][]{Lin94}, those usually lead to less straightforward methods for detection. For the rest of this paper, we will always mean by a "blob" of an image $f$ a local minimiser of $\normlap L$.

\removed{A detected blob $(\x, t)$ can be visualised by a circle with center $\x$ and radius proportional to $\sqrt{t}$. A common choice for the radius is $\sqrt{2 t}$, motivated by the fact that such a circle contains 68\% of the total mass of a Gaussian with center $\x$ and scale parameter $t$.}

Note that the scale-normalisation in \eqref{eq:normalized_laplacian} is necessary to achieve scale-invariance of the detected interest points: If a blob of scale $t$ is detected in the image $f$, then it should be detected as a blob of scale $s \cdot t$ in the rescaled image $\tilde f(\x) = f(\frac{\x}{s})$.

To build some intuition on the LoG method, consider the prototypical Gaussian blob with scale parameter $s$ and center $\bm m$:
\begin{displaymath}
f(\x) =  \frac{1}{2 \pi s} e^{-\frac{\norm{\x - \bm m}^2}{2 s}}.
\end{displaymath}
It can then be shown that the scale-normalised Laplacian $\normlap L$ has exactly one local minimum at the point $(\x, t) = (\bm m, s)$. Hence, we can recover the scale and center of the Gaussian blob by determining the local minimum of $\normlap L$ in the 3-dimensional scale space. Furthermore, this corresponds to the intuition that blobs of an image correspond to points where the intensity of an image has highly negative or positive curvature, corresponding to structures that are bright or dark, respectively.


For a practical implementation, the LoG method has to be formulated for discrete images $\f \in \R^{\removed{T \times }Z\added{ \times T}}$ and given a discrete set of scales $0 \leq t_1 < \ldots < t_K$. Analogously to the continuous setting, the scale-space representation of a discrete image $\f$ is obtained from a discrete convolution,
\begin{displaymath}
\bm L_t = \bm k_t * \f.
\end{displaymath}
As convolution kernel $\bm k_t$ a sampled Gaussian is often used. However, this choice is not consistent with the scale-space axioms \citep[see e.g. section 1.2.2. in][]{Wei98}, as discussed by \cite{Lin90}. Instead, the correct analogue of the continuous Gaussian kernel in the sense of scale-space theory has to be defined in terms of the modified Bessel function, i.e.
\begin{eqnarray}
k_{t, ij} & = & e^{-t^2} I_{i}(t) I_{j}(t), \label{eq:discrete_gaussian1}\\
I_i(t) & = & \sum_{\nu=0}^\infty \frac{1}{\nu! \Gamma(\nu + i + 1)} \left( \frac{t}{2} \right)^{2 \nu + i}. \label{eq:discrete_gaussian2}
\end{eqnarray}
With this choice, the scale-space properties from the continuous setting hold exactly in the discrete setting \citep{Lin91}. We have observed in our numerical tests that the use of this kernel leads to slightly better results than with the sampled Gaussian.

Given a discrete set of scales $0 \leq t_1 < \ldots < t_K$, we obtain a discrete scale-space representation of the image $\f$ as three-dimensional matrix $\bm L = (\bm L_{t_q})_{q=1}^K \in \R^{\removed{T \times }Z\added{ \times T} \times K}$. Similar to the continuous setting, blobs of the discrete image $\f$ are then detected as the local minimizers \removed{and maximizers} of $\disnormlap \bm L = (\disnormlap \bm L_{t_q})_{q=1}^K \in \R^{\removed{T \times }Z\added{ \times T} \times K}$, where $\disnormlap$ is the discrete scale-normalised Laplacian operator,
\begin{equation*}
\disnormlap \bm L_{ijt} = t \cdot (L_{i-1,j,t} + L_{i+1,j,t} + L_{i,j-1,t} + L_{i,j+1,t} - 4 L_{ijt}),
\end{equation*}
which is well-defined for all $(i,j) \in \lbrace 1, \ldots, \removed{T}\added{Z} \rbrace \times \lbrace 1, \ldots, \removed{Z}\added{T} \rbrace$ by reflecting the image at the boundary. \added{The numerical value of $\abs{\disnormlap \bm L}$ at a detected blob can be used as surrogate for the blob strength.}

\added{It is recommended to post-process the results of the LoG method by removing blobs that are too weak or overlap too much. First, all blobs for which the value of $\abs{\disnormlap \bm L}$ is below a certain threshold $l^\mathrm{tresh}$ are removed. In the present work, we use a relative threshold of the form $l^\mathrm{tresh} = r^\mathrm{tresh} \cdot \abs{\min \normlap \ssrmap}$, where $r^\mathrm{tresh} \in (0, 1)$. Without thresholding, the LoG method would detect blobs at the slightest fluctuations of the image intensity, which would
often not correspond to meaningful features. 

In a second step, the relative overlap of the detected blobs is computed. Hereby, a blob $\bm b = (i, j, q)$ is identified with a circle of radius $2 \sqrt{t_q}$, centered at the pixel $(i, j)$. If the relative overlap of two blobs surpasses a treshold $o^\mathrm{tresh}_1$, the blob with the smaller value of $\abs{\disnormlap \bm L}$ (i.e. the weaker blob) is removed.}

\added{We visualise a detected blob $(i, j, t)$ by a circle centered at the pixel $(i,j)$ and radius proportional to $\sqrt{t}$. A common choice for the radius is $\sqrt{2 t}$, motivated by the fact that such a circle contains 68\% of the total mass of a Gaussian with center $(i,j)$ and scale parameter $t$.}

\added{In figure \ref{fig:log_demo}, we plot the results of applying the LoG method to a MAP estimate of the stellar distribution function from mock data. In this particular case we used the relative threshold $r^\mathrm{thresh}=0.02$ for the blob strength and the threshold $o^\mathrm{thresh}=0.5$ for the maximal relative overlap of two blobs.} 

\added{Before we move on, we want to point out that while the post-processing step is important, it should not be interpreted as a statistical procedure since it does not take the measurement error into account but only depends on the given point estimate of the distribution function. For example, a blob in the MAP estimate might be weak (i.e. a low value of $\abs{\disnormlap \bm L}$) but its presence could be very certain. Vice versa, a very strong blob ($\abs{\disnormlap \bm L}$ large) in the estimate might at the same time be very uncertain.}

\section{An SVD-MCMC method for scalable uncertainty quantification}\label{sec:svd_mcmc}

\removed{
The ULoG method described in section \ref{subsec:ulog} depends on the posterior $p(\f | \obs)$ only through the filtered credible intervals. This means that for a practical implementation, it can be combined with any "backend" that yields estimates of such intervals, in particular standard methods such as MCMC or Variational Bayes. However, our present problem poses some distinct challenges due to the combination of ill-posedness, high-dimensionality and nonnegativity constraints, for which generic methods for uncertainty-quantification are less well-suited.
}
\added{
We wish to generalise the LoG method for blob detection to handle uncertainty. As a prerequiste to this, we first must explain how to quantify uncertainties for our problem. This poses some distinct challenges due to the combination of ill-posedness, high-dimensionality and nonnegativity constraints, for which generic methods for uncertainty quantification are not well-suited.
}

In this section, we present a specialised MCMC method that is able to overcome the computational challenges in the reconstruction of stellar distribution functions. To make MCMC sampling computationally feasible for this problem we employ the following strategy: reduce the dimensions of the problem, sample the posterior in the reduced space, then deproject the solution back to the original space. The specific dimensionality reduction technique we will use is based on truncating the singular value decomposition (SVD) which is closely related to principal component analysis (PCA). This basic idea follows \citet{ppca_99} however our method for deprojection is substantially different since we sample possible deprojections rather than take a single choice. In the context of stellar population modelling, previous work includes \cite{bayeslosvd} who also combine SVD and MCMC sampling. Here, we build on this work by outlining how to deproject samples from the low-dimensional space in order to perform inference on stellar populations.

\subsection{Notation}\label{sssec:mcmc_notation}

In this section it will be useful to re\removed{-}write our model in a vectorised form, i.e. where we reshape 2D matrices to 1D vectors, 3D tensors to 2D matrices etc. Rather than introduce new baroque notation, for the duration of Section \ref{sec:svd_mcmc} we ask the reader to re\removed{-}interpret existing notation in this fashion. Specifically, the foward model in \removed{Equation }\eqref{eq:y_F}, i.e.
\begin{equation}
\y = \G \f + \bm{\epsilon},
\label{eq:y_F_repeated}
\end{equation}
originally had $\G$ as a 3D tensor of shape $(n,\removed{T,}Z\added{, T})$ and $\f$ as the 2D image of shape $(\removed{T,}Z\added{,T})$. Here we abuse notation and instead use $\f$ to refer to the 1D vectorised image of length $\removed{TZ}\added{Z \cdot T}$ and $\G$ to be correspondingly reshaped into a 2D matrix of shape $(n,\removed{TZ}\added{Z \cdot T})$. Similarly the prior covariance $\Sigma_{\f}$ from equation \eqref{eq:truncated_gaussian_prior} should now be considered a 2D covariance matrix rather than a 4-tensor.

\subsection{SVD regression}\label{sssec:svd_regression}

The SVD is a matrix decomposition which generalises eigenvalue decomposition to rectangular matrices. For example, given a matrix $\G$ with shape $(n,p=\removed{T}Z\added{T})$, the SVD uniquely factorises $\G$ into three matrices,
\begin{equation*}
\G = \U \bm{S} \V^T.
\end{equation*}
The matrix $\bm{S}$ is a diagonal matrix whose entries are known as singular values, which are non-negative and are sorted in decreasing order, i.e. $\bm{S}[0,0]\geq\bm{S}[1,1]\geq...\geq 0$. These are analogous to eigenvalues of an eigendecomposition. The columns of $\U$ and $\V$ contain orthonormal basis-vectors of the spaces $\mathbb{R}^n$ and $\mathbb{R}^p$ respectively, which are analogous to eigenvectors.

The SVD can be used for dimensionality reduction if the singular values decay sufficiently fast. The \emph{truncated SVD} retains only the first $q$ components of the factorisation, i.e.
\begin{equation}
\U_q = \U[:, :q],\;\;
\bm{S}_q = \bm{S}[:q, :q],\;\;
\V_q = \V[:,:q]. \label{eqn:truncated_SVD}
\end{equation}
The total squared-error incurred by approximating $\G$ by the truncated SVD is given by the sum of squares of the remaining $p-q$ singular values, i.e.
\begin{equation*}
\norm{\G - \U_q \bm{S}_q \V_q^\top}_2^2 = \sum_{i=q+1}^{p} \bm{S}[i, i]^2.
\end{equation*}
Therefore, if the singular values decay sufficiently fast then the truncated SVD can provide an accurate approximation of $\G$ for relatively small values of $q$.
\removed{The top panel of Figure \ref{fig:svd} shows the singular values where the matrix $\G$ is set to the MILES SSP \citep{Vazdekis10} templates defined in Section \ref{ssec:mock_data}. The largest and smallest singular value of this matrix are separated by a factor of $\sim 10^8$ - a quantity known as the \emph{condition number} of the matrix - however they do not decay at a uniform rate. The steep initial drop in singular values means that very small choices of $q$ are able to provide accurate approximations to $\G$.}
\added{For the MILES SSP models \citep{Vazdekis10} we find that $q=15$ provides a sufficiently good approximation. In general this choice will depend on the particular SSP models used and the data quality, and} in \removed{section} \added{appendix} \ref{sssec:svd_choose_q} we suggest a general strategy to choose $q$.

The truncated SVD transforms our problem to a low-dimensional regression problem. We introduce the notation
\begin{equation}
\Z = \U_q \bm{S}_q
\label{eqn:Z_H}
\end{equation}
and insert into equation \eqref{eq:y_F_repeated} to give the transformed equation for the forward model,
\begin{align}
&\y \approx \Z \bm{\eta} + \bm{\epsilon} \;\; \text{where}
\label{eq:y_Zeta} \\
&\bm{\eta} = \V_q^\top \f.
\label{eq:eta}
\end{align}
The original $p$-dimensional parameter vector $\f$ in equation \eqref{eq:y_F_repeated} has been reduced to a $q$-dimensional vector $\bm{\eta}$ in equation \eqref{eq:y_Zeta}. To do inference on $\f$, we first do inference on $\bm{\eta}$ using equation \eqref{eq:y_Zeta}, then deproject back to infer $\f$. Having reduced the dimensions and the condition-number of the problem, this problem is now computationally more amenable to MCMC sampling to estimate the posterior distribution $p(\bm{\eta}|\y)$.

\subsection{Deprojection}\label{sssec:svd_deprojection}

How should we perform the deprojection? The suggestion from \citet{ppca_99} is to use the least-square deprojection of equation \eqref{eq:eta}, i.e. for each sample $\bm{\eta}$ from the MCMC, calculate
\begin{equation*}
\f_* = \V_q \bm{\eta}.
\end{equation*}
This satisfies equation \eqref{eq:eta} because the SVD ensures that $\V_q$ is orthonormal, i.e. $\V_q \V_q^\top = \I_q$. \emph{The least-square deprojection is not suitable for our problem, however, since it is not guaranteed to satisfy the non-negativity constraint on $\f$}. In numerical tests, these deprojections did contain many negative entries and are therefore \removed{unacceptable on physical grounds}\added{represent unphysical solutions}. A solution to this problem would be to instead perform a constrained deprojection, where we minimise some function $\mathcal{R}(\f)$ subject to the desired constraints,
\begin{equation}
\f_*  = \argmin_{\f} \; \mathcal{R}(\f) \;
      \text{subject to} \; \V_q^\top \f = \bm{\eta} \;
      \text{and} \; \f \geq 0.
\label{eq:constrained_deproj}
\end{equation}
A natural choice for the function $\mathcal{R}(\f)$ is to use the same regularising penalty function defined in equation \eqref{eq:map}. For standard, quadratic choices of $\mathcal{R}(\f)$, equation \eqref{eq:constrained_deproj} is a quadratic programming problem which can be solved efficiently. In practice, however, we find that this solution is unacceptable since these deprojections are unphysically spiky.
Although the objective function in equation \eqref{eq:constrained_deproj} promotes smoothness in the deprojections, the equality constraint $\V_q^\top \f = \bm{\eta}$ \emph{overwhelms} this effect, resulting in unphysical deprojections too similar to the spiky ML solution.

The final strategy we investigated is to again use MCMC sampling. Whereas deprojection by calculating a point estimate will neglect uncertainty in the $p-q$ SVD components which are not informed by the data, by sampling we can account for this. These components will instead be informed by the prior $p(\f)$ defined in equation \eqref{eq:prior}, i.e. the Bayesian equivalent of the regularisation term. Since the strength of this prior depends on a tunable regularisation parameter $\beta$, samples deprojected in this way can be made as smooth as desired, in analogy with MAP vs ML recoveries.

\subsection{SVD vs. PCA}\label{sssec:svd_vs_pca}

SVD and PCA are closely related techniques. If we consider the matrix $\G$ to be a data matrix with $n$ samples each of dimension $p$, and then we column\added{-}center this matrix, i.e.
\begin{align}
&\hat{\G} = \G - [\vmu, ..., \vmu]  \;\;  \mathrm{where} \label{eqn:center_G} \\
&\vmu = \frac{1}{p} \sum_{i=1}^{p} \g_i
\end{align}
and $\g_i$ are the columns of $\G$, then the SVD of $\hat{\G}$,
\begin{equation*}
\hat{\G} = \hat{\U} \hat{\bm{S}} \hat{\V}^\top,
\end{equation*}
is equivalent to PCA of the data. They are equivalent in the sense that the matrix $\hat{\V}$ contains the principal axes of the data and the squares of the singular values in $\hat{\bm{S}}$ acquire the interpretation of being the variances explained by each principal component. In practice we find that using PCA works better than performing an SVD directly on $\G$, so we adopt this choice throughout this work. We retain the name \emph{SVD-MCMC} since the SVD interpretation helps to clarify the underlying ideas, and to prevent confusion with the more common usage of PCA as applied to observed spectra rather than model templates \citep[e.g.][]{Yip04}.

Centering $\G$ as in equation \eqref{eqn:center_G} transforms the forward model in \eqref{eq:y_F_repeated} as follows,
\begin{equation*}
\y = m \vmu + \hat{\G} \f + \bm{\epsilon},
\end{equation*}
where $m$ is a scalar constant subject to the constraint
\begin{equation*}
m = \sum_{i=1}^{p} f_i.
\end{equation*}
We define the truncated SVD of $\hat{\G}$ equivalently to equation \eqref{eqn:truncated_SVD} to give $\hat{\U_q}, \hat{\bm{S}}_q$ and $\hat{\V_q}$, then define $\hat{\Z}$ equivalently to equation \eqref{eqn:Z_H} to give the reduced forward model after PCA,
\begin{align*}
&\y \approx m \vmu + \hat{\Z} \bm{\eta} + \bm{\epsilon}
\;\; \text{where} \\
&\bm{\eta} = \hat{\V}^\top \f.
\end{align*}

\subsection{Probabilistic model and MCMC implementation}\label{sssec:mcmc_implementation}

We can now define our probabilistic models and describe how we sample their posterior distributions. A probabilistic model is defined as a series of sampling statements and transformations of variables required to generate the observed data. Combined with Bayes' theorem, the model implicitly defines an expression for the posterior probability function of the model unknowns given the observed data. An MCMC algorithm evaluates this expression to draw samples from the posterior.

We will compare two models, the \emph{Full MCMC} model with no dimensionality reduction, and the  \emph{SVD-MCMC} model using the truncated SVD. Both models start by sampling the prior distribution on $\f$, i.e.
\begin{equation}
	p(\f) \propto \begin{cases}
		\mathcal{N}\left(\0, \removed{- \frac{\beta}{2}}\added{\frac{1}{\beta}} \Covf\right), & \text{if } \f \geq 0, \label{eqn:f_prior_probmod} \\
		0, & \text{otherwise}.
		\end{cases}
\end{equation}
The two models are:
\begin{enumerate}
	\item Full MCMC:
		\begin{align*}
		\f & \sim p(\f)\\
		\y \added{~|~ \f} & \sim \mathcal{N}(\G \f , \Noicov).
		\end{align*}
		\item SVD-MCMC (for some choice of $q$):
			\begin{align*}
			\f &\sim p(\f)\\
			\bm{\eta} &= \hat{\V}_q^\top \f, \\
			m &= \sum_{i=1}^{p} f_i, \\
			\y \added{~|~ \bm \eta, m} & \sim \mathcal{N}(m \vmu + \hat{\Z} \bm{\eta} , \Noicov).
			\end{align*}
\end{enumerate}
In both cases, the final sampling statement draws data $\y$ from a multivariate normal with noise covariance $\Noicov$ as defined in equation \eqref{eq:noise}.
Each model implicitly defines an expression for the posterior density $p(\f|\y)$, which we sample via MCMC.

For MCMC sampling, we use the Hamiltonian Monte Carlo algorithm \citep[HMC,][]{neal11} as implemented in the \texttt{numpyro} probabilistic programming language \citep{numpyro}. Unless stated otherwise, for all experiments shown we run a single chain for 5000 warmup steps and 10,000 sampling steps. The warmup parameters are left at the \texttt{numpyro} defaults. We check convergence with three criteria: (i) all sampled parameters have an $\hat{R}$ statistic \citep{GelmanRubin92} in the range $[0.95,1.05]$, (ii) all sampled parameters have effective sample size \citep{Geyer20111IT} of at least 100, and (iii) no divergent transitions occur during sampling \citep{Betancourt16}.

Note that for the truncated SVD model, we directly sample the posterior $p(\f|\y)$, i.e. the \emph{deprojection} occurs internally during sampling. We did also try the alternative implementation of first sampling $p(\bm{\eta},m|\y)$ and then later deprojecting as a subsequent step.
The inferred posteriors and total computation times of the two implementations are very similar, however the first step of the alternative is orders of magnitude faster than the deprojection step.
Modelling with SVD-truncated spectral templates is therefore an appealing option for purely kinematic models, i.e. where stellar populations can be considered as a nuisance effect and need not be deprojected.
This is the approach taken by \cite{bayeslosvd}.
If we do also wish to also perform inference over stellar populations, however, we see no clear benefit in separating the two sampling steps.
Indeed, one possible downside to their separation is that sampling $p(\bm{\eta},m|\y)$ requires defining some arbitrary prior $p(\bm{\eta},m)$.
It is not possible for any such low-dimensional prior to be consistent with \eqref{eqn:f_prior_probmod}, or any physical prior, since the non-negativity constraint can only be satisfied in $p$-dimensions.
Nevertheless, we note that both implementations of SVD-MCMC enjoys a significant speed-up compared to the full model.

\removed{
\subsection{Computing filtered credible intervals from samples}\label{subsec:fcis_from_samples}

Given $S$ samples $\f^1, \ldots, \f^S \in \R^{T \times Z}$ from SVD-MCMC, a straightforward estimator for posterior filtered $(1-\alpha)$-credible intervals is the minimal-volume axis-aligned box that contains at least $(1-\alpha) \cdot S$ of the samples after convolution with the corresponding filter. The problem of computing such a box is known as the $k$-enclosing box problem in computational geometry, and there exist several algorithms for solving it exactly or approximately (see \cite{SegKed98} and \cite{ChaHar21}). However, the computational complexity of these algorithms increases at least quadratically with the number of samples and exponentially with the underlying dimension, which renders it impractical for our high-dimensional problem.
We also considered a classic method by \cite{BesGreHigMen95} for estimating simultaneous credible intervals, which can easily be modified to estimate \emph{filtered} credible intervals aswell. However, in our numerical tests this method often led to credible intervals that contained 100\% of the samples, rendering them useless. Instead, the following heuristic approach worked better: First, we convolved each sample $\f^i$ with a Gaussian kernel $\bm k_t$ corresponding to the desired scale. From the resulting filtered samples $\bm L_t^i = \bm k_t * \bm f^i$, we estimated \emph{marginal} filtered credible intervals using order statistics, yielding a multi-dimensional interval $[\bm \ell, \bm u] \subset \R^{T \times Z}$ (see equation \eqref{eq:multidimensional_interval}). Since $[\bm \ell, \bm u]$ usually contained too few samples (see the discussion in section \ref{subsec:credible_intervals}), it was then rescaled until it enclosed exactly $\lceil (1 - \alpha) \cdot S \rceil$ filtered samples. The rescaled multi-dimensional interval was then used as estimate for the filtered credible interval $[\Fcilow_t, \Fciupp_t]$.}

\section{Uncertainty-aware blob detection}\label{sec:uncertainty_aware_blob_detection}

\removed{In figure (\ref{fig:log_demo}), we visualise the results of applying the LoG method to a MAP estimate of the stellar distribution function from mock data.}
\added{How can we incorporate uncertainties into the LoG method for blob detection? Our motivation for doing so is demonsrated in figure \ref{fig:log_demo}: here we see that the LoG} \removed{The} method identifies multiple blobs in the \added{MAP estimate of the} distribution function. However, the MAP estimate depends on the measurement noise, and so it is uncertain whether structures that are identified in this way also exist in the true, unknown stellar distribution function. Since this question needs to be addressed if one wants to make credible statements about the structure of the stellar distribution function, one has to find a way to incorporate the a-posteriori uncertainty of the stellar distribution function into the blob detection method.

In this section, we show how this can be achieved by combining the ideas from scale-space theory with the concept of multi-dimensional Bayesian credible intervals, a central tool for understanding and visualising uncertainty in imaging. This results in an uncertainty-aware variant of the Laplacian-of-Gaussians (LoG) method, which we correspondingly refer to as \emph{ULoG}.

In section \ref{subsec:credible_intervals}, we will review Bayesian posterior credible intervals for images and related notions. In section \ref{subsec:filtered_credible_intervals}, we introduce a new type of credible intervals that incorporate\added{s} the notion of scale into the quantification of posterior uncertainty. We will refer to these as \emph{filtered credible intervals}. \added{Section \ref{subsec:fcis_from_samples} discusses how they can be estimated from MCMC samples.} \removed{Building on this idea, w}\added{W}e then describe how filtered credible intervals can be used to define an uncertainty-aware variant of the Laplacian-of-Gaussians method, ULoG. The details of this method are given in section \ref{subsec:ulog}.

\subsection{Credible intervals for imaging}\label{subsec:credible_intervals}

A central tool for Bayesian uncertainty quantification are \emph{posterior credible intervals}. In general, they do not coincide with frequentist \emph{confidence intervals} and should be interpreted differently. Since credible intervals are canonically only defined for scalar parameters, the definition has to be adapted when the unknown quantity of interest is multi-dimensional, as in the case of a discrete image.


\removed{
One such generalisation is the concept of a \emph{posterior credible region}. Given a posterior density function $p(\f | \obs)$ and a small credibility parameter $\alpha \in (0,1)$ (typical values are $\alpha=0.05$ and $\alpha=0.01$, respectively representing 95\%- and 99\%-credibility), we call a set $\region \subset \R^{T \times Z}$ a posterior $(1-\alpha)$-credible region for $\f$ if
\begin{displaymath}
\int_{\region} p(\f | \obs) \dif \f \geq 1 - \alpha.
\end{displaymath}


The fact that a credible region associated to an image $\f$ is a subset of $\R^{T \times Z}$ with a potentially nontrivial geometry naturally poses the question how it can be visualised. A typical solution is to constrain oneself to a certain subclass of credible regions that is easier to visualise, namely \emph{simultaneous credible intervals}.}

Before we state the definition, it will be useful to introduce the following notation for multi-dimensional intervals: Let $\bm \ell, \bm u \in \R^{\removed{T \times }Z\added{ \times T}}$ be two images such that $\ell_{ij} \leq u_{ij}$ for every pixel $(i, j)$. Then, we will refer to the hyper-box
\begin{equation}
[\bm \ell, \bm u] := \Set{\f \in \R^{\removed{T \times }Z\added{ \times T}}}{\bm \ell \leq \f \leq \bm u} \subset \R^{\removed{T \times }Z\added{ \times T}} \label{eq:multidimensional_interval}
\end{equation}
as a \emph{multi-dimensional interval} in $\R^{\removed{T \times }Z\added{ \times T}}$. \removed{With this definition,} \added{Given a posterior density function $p(\f | \obs)$ and a small credibility parameter $\alpha \in (0,1)$ (typical values are $\alpha=0.05$ and $\alpha=0.01$, respectively representing 95\%- and 99\%-credibility),} we \added{then }call a set $[\flow, \fupp] \subset \R^{\removed{T \times }Z\added{ \times T}}$ a simultaneous $(1-\alpha)$-credible interval if
\begin{equation}
P(\f \in [\flow, \fupp] | \obs) \geq 1 - \alpha. \label{eq:simultaneous_credible_intervals}
\end{equation}
\added{That is, the multi-dimensional interval $[\flow, \fupp]$ contains at least a given percentage of the posterior probability mass.} By looking at the plots of both the lower bound $\flow \in \R^{\removed{T \times }Z\added{ \times T}}$ and the upper bound $\fupp \in \R^{\removed{T \times }Z\added{ \times T}}$, \removed{a practitioner}\added{we} can obtain a visual understanding of the posterior uncertainty. \added{This is examplified in the second row of figure \ref{fig:credible_intervals}. The first row shows a simulated ground truth from which mock data was generated (see section \ref{ssec:mock_data}). The second row shows the corresponding MAP estimate $\fmapvec$, together with the lower bound $\flow$ and upper bound $\fupp$ of simultaneous 95\%-credible intervals.}

\removed{
The \emph{simultaneous} credible intervals as defined above need to be distinguished from \emph{marginal} credible intervals that are also often used for uncertainty visualisation (in one dimension, these notions coincide). Given a pixel $(i, j)$, we call an interval $[\ell_{ij}, u_{ij}] \subset \R$ a \emph{marginal posterior $(1-\alpha)$-credible interval} if it is a $(1-\alpha)$-credible interval with respect to the \emph{marginalised} posterior distribution $p(f_{ij} | \obs)$. That is, if there holds
\begin{displaymath}
P(f_{ij} \in [\ell_{ij}, u_{ij}] | \obs) \geq 1 - \alpha.
\end{displaymath}
If we assemble all lower bounds $(\ell_{ij})$ and upper bounds $(u_{ij})$ into images $\bm \ell, \bm u \in \R^{T \times Z}$, these images define then a multi-dimensional interval $[\bm \ell, \bm u] \subset \R^{T \times Z}$ via equation \eqref{eq:multidimensional_interval}. However, this interval will in general not define a simultaneous credible interval, as pointed out e. g. by \cite{Gan09}. For our purposes, simultaneous credible intervals are more suitable, since they allow the intuitive interpretation that the signal lies "between" the lower bound $\flow$ and the upper bound $\fupp$ with posterior probability greater or equal $1 - \alpha$. In other words, a simultaneous $(1-\alpha)$-credible interval is always a $(1-\alpha)$-credible region, while this might fail for a marginal $(1-\alpha)$-credible interval.

We also remark that the terminology in this area is not completely standardised. Credible regions and credible intervals are sometimes called Bayesian confidence regions and Bayesian confidence intervals, respectively, while the multi-dimensional simultaneous credible intervals defined above have been referred to as Bayesian confidence bands or credible bands in the area of time series analysis (see \cite{Bre78}).

In the rest of this paper, we will for the sake of brevity simply write "credible interval" when we a mean (multi-dimensional) "simultaneous credible interval".
}
\added{Note that the simultaneous credible intervals defined in equation \eqref{eq:simultaneous_credible_intervals} have to be interpreted differently than \emph{pixel-wise} credible intervals which also appear in the statistical literature. We discuss these differences and the motivation behind our choice in appendix \ref{app:simultaneous_vs_pixelwise}.}

\added{In section \ref{subsec:fcis_from_samples}, we describe how we compute simultaneous credible intervals using MCMC samples.}

\subsection{Filtered credible intervals}\label{subsec:filtered_credible_intervals}

\begin{figure*}
\resizebox{\hsize}{!}{\includegraphics{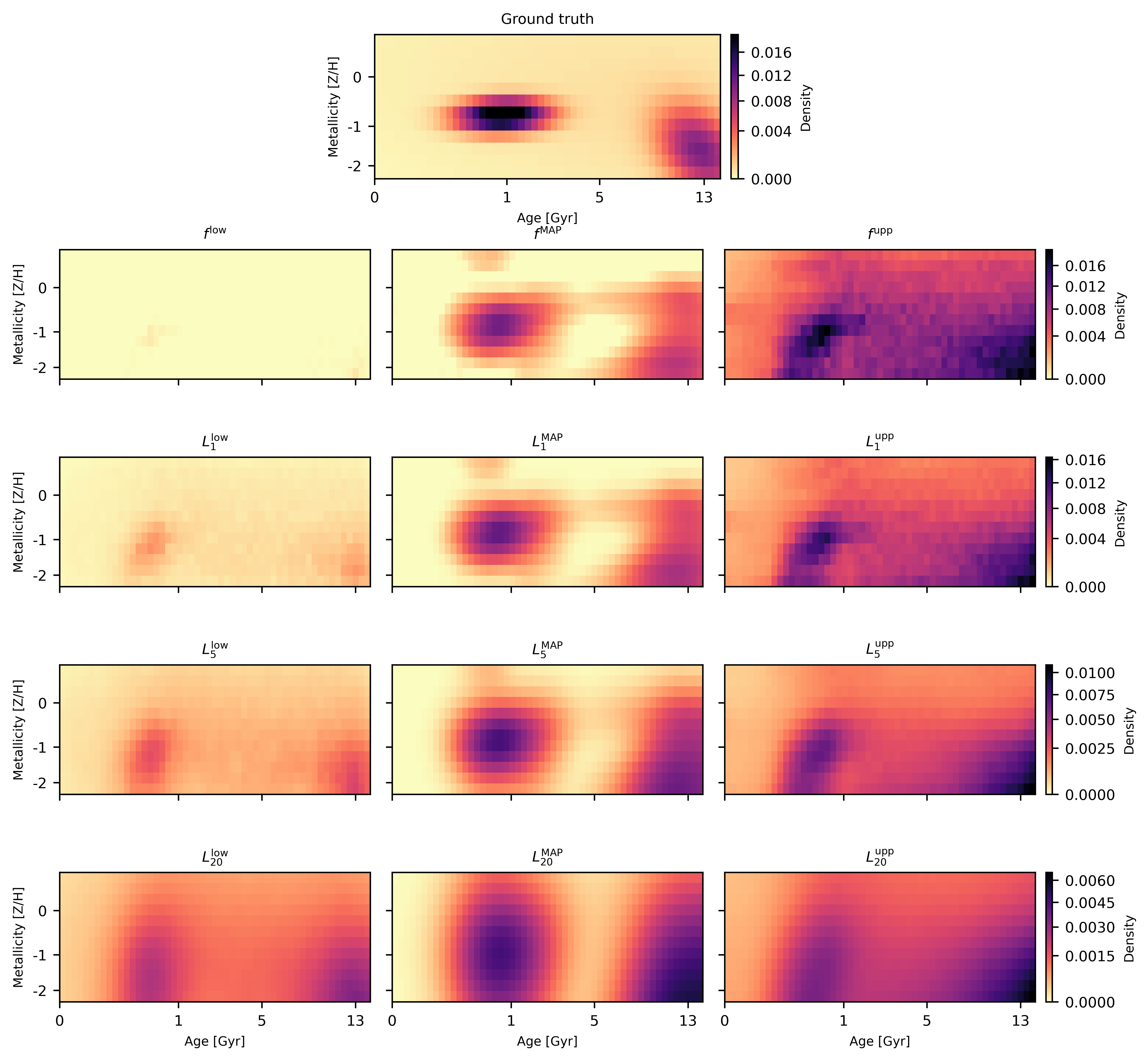}}
      \caption{Visualisation of credible intervals from mock data. The upper panel shows the ground truth from which the mock data was generated. In the second row, the MAP estimate $\fmapvec$ (center column) together with the lower bound $\flow$ (left) and upper bound $\fupp$ (right) of the \removed{pixel-wise}\added{(unfiltered) simultaneous} credible intervals are plotted. Below that are the lower and upper bound of filtered credible intervals at 3 different scales, together with the corresponding filtered MAP estimate $\bm L_t^\mathrm{MAP} = \bm k_t * \fmapvec$.}
\label{fig:credible_intervals}
\end{figure*}

As a naive approach to perform uncertainty-aware blob detection we might try to identify the blobs directly from the \removed{pixel-wise}\added{simultaneous} credible intervals. However, if the image of interest is highly uncertain, as is the case in the present problem of stellar recoveries, such an approach is rather limited. This can be seen from the first two rows of figure \ref{fig:credible_intervals}. \removed{The first row shows a simulated ground truth from which mock data was generated (see section \ref{ssec:mock_data}). The second row shows the corresponding MAP estimate $\fmapvec$, together with the lower bound $\flow$ and upper bound $\fupp$ of pixel-wise 95\%-credible intervals.} One can observe that the MAP estimate is quite close to the ground truth. In contrast, the \removed{pixel-wise }credible intervals are so wide that they only allow for very rough statements on the shape of the stellar distribution function. The lower bound is close to zero for most pixels. Clearly, the \removed{pixel-wise }credible intervals alone do not accurately represent the quality of our estimate.

This problem can be resolved by recognising that we are less interested in the intensity values of specific pixels, but in higher-scale \emph{structures}. Hence, we are actually interested in coarse-scale information. It is an important insight that -- in contrast to the case where the image of interest is treated as given -- the uncertainty with respect to higher scales cannot be directly derived from the uncertainties on finer scales. The correlation structure in the posterior distribution might be more complex and exhibit long-range dependencies, which means that the overall uncertainty cannot be reduced to the pixel-by-pixel case.

Consequently, if we want more informative uncertainty quantification, we have to integrate the notion of \emph{scale}. This is achieved in our work by integrating the ideas of scale-space representations (see also section \ref{subsec:scale_space}): Instead of computing credible intervals for the vector of pixel intensities $\f$, we can perform uncertainty quantification subject to a scale parameter $t$ by computing credible intervals for the smoothed image $\bm L_t = \bm k_t * \f$, where $\bm k_t$ is the discrete analogue of the Gaussian kernel, defined in equations \eqref{eq:discrete_gaussian1}-\eqref{eq:discrete_gaussian2}. That is, we compute images $\Fcilow_t, \Fciupp_t \in \R^{\removed{T \times }Z\added{ \times T}}$ such that
\begin{displaymath}
P(\bm L_t \in [\Fcilow_t, \Fciupp_t] | \obs) \geq 1 - \alpha.
\end{displaymath}
We will refer to $[\Fcilow_t, \Fciupp_t]$ as a \emph{filtered credible interval at scale $t$}.

In the last three rows of figure \ref{fig:credible_intervals}, filtered credible intervals at three different scales are plotted. We see that the intervals tighten with increasing scale, but loose the ability to represent small structures. This visualises a very intuitive trade-off between scale and uncertainty: The uncertainty is lower with respect to larger (high-scale) structures, since they have less degrees of freedom. Correspondingly, the uncertainty with respect to small-scale structures is larger, as the given image has much more degrees of freedom at the small scale. The advantage of filtered credible intervals is that they allow for an explicit choice. If one is only interested in high-scale information, they allow to remove the superfluous uncertainty with respect to small-scale variations. If one moves to a finer scale, one has to pay the price of potentially larger uncertainty.

\added{
\subsection{\removed{Computing filtered} \added{Estimating} credible intervals from samples}\label{subsec:fcis_from_samples}

Since in our case there are no analytical expressions for either unfiltered or filtered credible intervals, we have to estimate both using MCMC samples. After generating $S$ samples $\f^1, \ldots, \f^S \in \R^{Z \times T}$ from the posterior distribution $p(\f | \obs)$, we estimate simultaneous $(1-\alpha)$-credible intervals $[\flow, \fupp]$ by computing a multi-dimensional interval that contains the desired percentage of samples $\f^i$. That is, we find $\flow  \in \R^{Z \times T}$ and $\fupp \in \R^{Z \times T}$ such that at least $(1-\alpha) \cdot S$ samples satisfy
\begin{align*}
\f^s \in [\flow, \fupp].
\end{align*}
Similarly, we estimate filtered credible intervals, at scale $t$, by computing a multi-dimensional interval that contains the desired percentage of \emph{filtered} samples. That is, we find $\Fcilow_t, \Fciupp_t \in \R^{Z \times T}$ such that at least $(1-\alpha) \cdot S$ samples satisfy
\begin{align*}
 \bm k_t * \f^s \in [\Fcilow_t, \Fciupp_t].
\end{align*}
In both cases, the estimation of credible intervals reduces to the following computational task: Given a sample of images $\bm x^1, \ldots, \bm x^S \in \R^{Z \times T}$, find the smallest multi-dimensional interval $[\bm \ell, \bm u] \subset \R^{Z \times T}$ that contains at least $(1 - \alpha)\cdot S$ of these samples (with $\bm x^s = \f^s$, this procedure estimates simultaneous credible intervals; with $\bm x^s = \bm k_t * \f^s$, this procedure estimates filtered credible intervals at scale $t$). This task is known as the $k$-enclosing box problem in computational geometry. We use the following heuristic but fast approach: First, find $\bm \ell, \bm u \in \R^{Z \times T}$ such that for every pixel $(i,j)$, there are at least $(1-\alpha)\cdot S$ images for which
\begin{align*}
\ell_{ij} \leq x^s_{ij} \leq u_{ij}
\end{align*}
(this is done using the order statistics of $(x^s_{ij})_{s=1}^S$). Since the resulting multi-dimensional interval $[\bm \ell, \bm u]$ usually contains too few points, it is then rescaled until it encloses at least $(1 - \alpha) \cdot S$ points.

Note that there are several existing algorithms for solving the $k$-enclosing box problem exactly or approximately (see \cite{SegKed98} and \cite{ChaHar21}). However, the computational complexity of these algorithms increases at least quadratically with the number of samples and exponentially with the underlying dimension, which renders them impractical for our high-dimensional problem.
We also considered a classic method by \cite{BesGreHigMen95}. However, in our numerical tests this method often led to credible intervals that contained 100\% of the samples, rendering them useless. For our purposes, the heuristic approach was sufficient.
}

\subsection{Uncertainty-aware Laplacian-of-Gaussians (ULoG)}\label{subsec:ulog}

\begin{figure}
   \centering
   \includegraphics[width=\hsize]{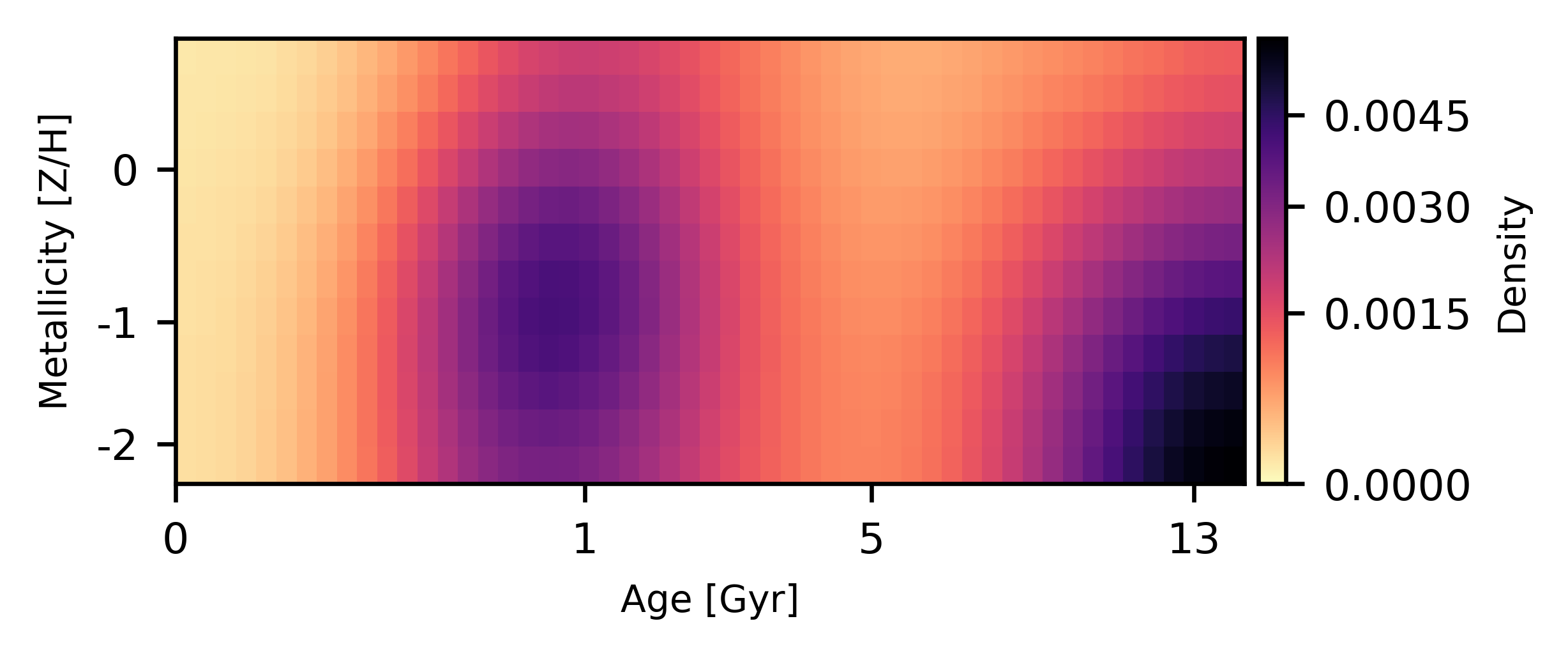}
      \caption{\added{Image of a $t$-blanket $\blanket_t$ ($t = 7$). The mock data is the same as in figures \ref{fig:log_demo} and \ref{fig:significant_blobs}. Two main structures are visible which correspond to the two significant blobs visualised in figure \ref{fig:significant_blobs}.}}
         \label{fig:blanket}
\end{figure}

In contrast to \added{simultaneous} credible intervals on the level of individual pixels, the filtered credible intervals in figure \ref{fig:credible_intervals} show that there are some structures that are "significant", in the sense that they are visible both in the image of the lower and the upper bound. This notion of significance is scale-dependent, since some structures might only be visible in the filtered credible intervals at sufficiently large scales. Intuitively, we would like to say that a blob is "credible" (at a given scale $t$) if all images $\bm h \in [\Fcilow_t, \Fciupp_t]$ exhibit a matching blob. Of course, one has to find a meaningful definition of what "matching" means, as the exact position and size of detectable blobs will usually vary among the images in $[\Fcilow_t, \Fciupp_t]$. Since checking all these images is computationally infeasible, we instead define a suitable representative $\blanket_t \in [\Fcilow_t, \Fciupp_t]$ that has "the least amount of blobs" in some quantitative sense. There are many different loss functions that penalise the amount of blobs in an image. We experimented with various loss functions of the form $J(\bm h) = \norm{\bm D \bm h}^2$ with $\bm D$ being a discrete differential operator. Among all the differential operators we tested, the discrete Laplacian $\bm \Delta$ led to the most favorable properties. We hence consider the representative $\blanket_t \in [\Fcilow_t, \Fciupp_t]$ given by the solution to the optimisation problem
\begin{equation}
\blanket_t = \argmin_{\bm h \in \R^{\removed{T \times }Z\added{ \times T}}} \Set{\norm{\bm \Delta \bm h}_2^2}{\Fcilow_t \leq \bm h \leq \Fciupp_t}. \label{eq:blanket}
\end{equation}
This definition can be further motivated by the fact that in the LoG method, blobs are related to the local minima and maxima of the discrete Laplacian of an image (see section \ref{subsec:laplacian_of_gaussians}). Hence, penalising $\norm{\bm \Delta \blanket_t}$ dampens the Laplacian and yields a feature-poor representative. On an intuitive level, solving \eqref{eq:blanket} can be thought of as spanning a two-dimensional "blanket" between the lower bound $\Fcilow_t$ and the upper bound $\Fciupp_t$. In accordance with this intuition, we will refer to the resulting image $\blanket_t$ as a \emph{$t$-blanket}. \added{An image of a $t$-blanket generated from mock data is shown in figure \ref{fig:blanket}.}

\begin{figure}
   \centering
   \includegraphics[width=\hsize]{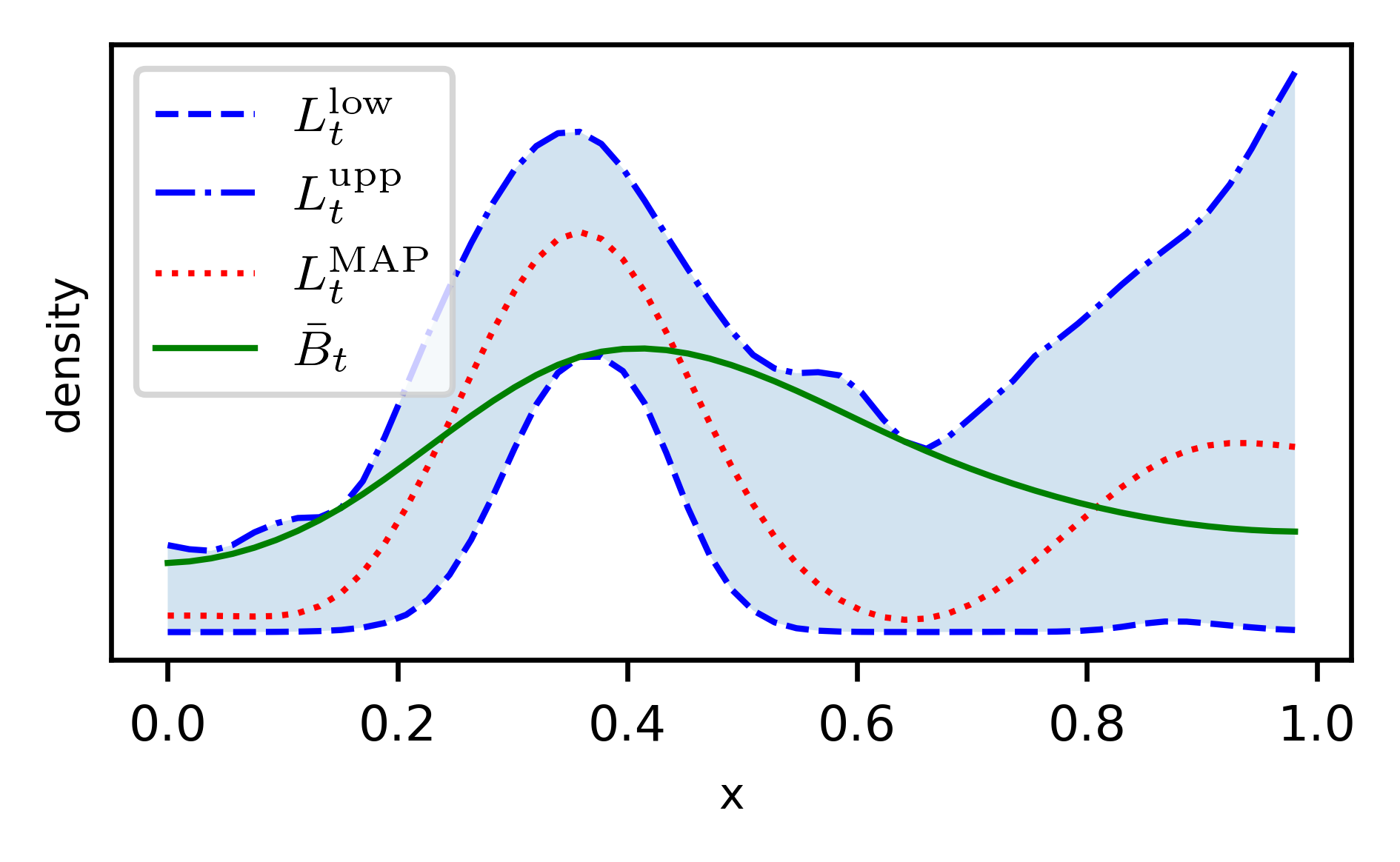}
      \caption{Illustration of the idea behind definition \eqref{eq:blanket}. Given a filtered credible interval (shaded region) determined by a lower bound $\fcilow_t$ and an upper bound $\fciupp_t$, we find a function $\blanket_t$ between these bounds that minimises the number of detectable blobs. The dotted, red graph $L^\mathrm{MAP}_t$ is the filtered MAP estimate. This is shown for some dummy variable $x$.}
         \label{fig:one_dimensional_example}
\end{figure}

To illustrate the intuition behind definition \eqref{eq:blanket}, a one-dimensional, continuous example is shown in figure \ref{fig:one_dimensional_example}. Given a MAP estimate $f^\mathrm{MAP}: [0, 1] \to \R$ of an unknown signal, assume that a posterior filtered credible interval is spanned by two functions $\fcilow_t: [0, 1] \to \R$ and $\fciupp_t: [0,1] \to \R$. The $t$-blanket $\bar B_t: [0, 1] \to \R$ is then the minimiser of $\norm{h''}$ over all functions $h$ that satisfy the pointwise constraint $\fcilow_t(x) \leq h(x) \leq \fciupp_t(x)$ and the boundary conditions $h'(0) = h'(1) = 0$. The blanket $\bar B_t$ should not be confused with the filtered MAP estimate $L^\mathrm{MAP}_t$, which will in general be less smooth and might exhibit more blobs than $\bar B_t$, as seen in figure \ref{fig:one_dimensional_example}.

If we do not treat the scale $t$ as given, we can study the three-dimensional scale-space structure of the "stack" of blankets $\blanket = (\blanket_t)_{t \geq 0} \subset \R^{\removed{T \times }Z\added{ \times T}}$. We observed in numerical tests that any given blob will typically be visible in $\blanket_t$ for a range of scales $[t_1, t_2] \subset [0, \infty)$. At the lowest scale $t_1$, it will typically be very weak, then grow stronger as the scale increases (and the uncertainty decreases, correspondingly) until it reaches a peak strength at some scale $t^*$. Then, as the scale increases further towards $t_2$, the strength of the blob decreases again as it is "smoothed out". This phenomenon is somewhat analogous to the deterministic theory of scale-space blobs \citep[see][]{Lin91}. Motivated by this similarity, we consider the blanket stack $\blanket$ as an "uncertainty-aware" scale-space representation. In a way, it represents our posterior uncertainty with respect to the unknown image $\f$ in scale space. In analogy to the LoG method from section \ref{subsec:laplacian_of_gaussians}, we thus propose the heuristic principle that "significant" scale-space blobs correspond to local minimisers of $\normlap \blanket$ in scale-space.

\added{
Instead of considering these significant blobs alone, we can obtain a more nuanced view of our information on the unknown stellar distribution by matching the set of candidate blobs detected in the MAP estimate $\fmap$ (see e.g. figure \ref{fig:log_demo}) to the significant blobs detected from the blanket stack $\blanket$. For example, if a blob is detected in the MAP but does not correspond to a significant blob in $\blanket$, it does not mean that it can be ignored, but only that its presence is too uncertain with respect to the given credibility level $1 -\alpha$.}

\removed{Based on this principle}\added{In summary}, we arrive at \removed{an}\added{the following} uncertainty-aware variant of the LoG method, which we refer to as ULoG. Given a discrete sequence of scales $0 \leq t_1 <  \ldots < t_K$, do the following:

\begin{enumerate}
\item Perform blob detection on the MAP estimate $\fmapvec$ using the LoG method described in section \ref{subsec:laplacian_of_gaussians}. Store the detected blobs in a set $\mathcal B \subset \removed{\range{T} \times }\range{Z}\added{ \times \range{T}} \times \lbrace 1, \ldots, K \rbrace$.
\item For $k=1,\ldots,K$, do:
\begin{enumerate}
\item Compute a filtered credible interval $[\Fcilow_{t_q}, \Fciupp_{t_q}]$.
\item Using $[\Fcilow_{t_q}, \Fciupp_{t_q}]$, compute the $t_q$-blanket $\blanket_{t_q}$ by solving \eqref{eq:blanket}.
\end{enumerate}
\item Stack the blankets in a 3-dimensional array $\blanket = (\blanket_{t_q})_{q=1}^K$.
\item Determine the local minimisers of $\bm \normlap \blanket \in \R^{\removed{T \times }Z\added{ \times T} \times K}$ and store them in a set $\mathcal C$.
\item Perform a matching procedure (see section \ref{sssec:matching}) that matches elements of $\mathcal B$ to elements of $\mathcal C$.\removed{ If a blob $\bm b \in \mathcal B$ has a matching element $\bm c \in \mathcal C$, it is marked as significant. If there is no match, the blob $\bm b$ is marked as "not significant".}
\item \added{Visualise the results (see section \ref{sssec:visualise}).} 
\end{enumerate}

\removed{
There are multiple ways in which we could visualise the results of the outlined algorithm. In figure \ref{fig:significant_blobs}, the following procedure was applied:
\begin{itemize}
\item A scale-space blob $\bm b = (i, j, q)$ is visualised as a circle with center $(i, j)$ and radius $\sqrt{2 t_q}$ (cf. section \ref{subsec:laplacian_of_gaussians}).
\item If a blob $\bm b \in \mathcal B$ is "not significant", it is visualised as a red circle.
\item If a blob $\bm b \in \mathcal B$ is "significant", i.e. if there is a matching blob $\bm c \in \mathcal C$, it is visualised as a green circle. Furthermore, the match $\bm c$ is visualised as a dashed, blue circle.
\end{itemize}

The dashed blue circles have to be interpreted in the following way: If there is a second structure inside the circle, then we cannot distinguish it with confidence from the corresponding blob. Hence, the radius of the circle indicates how well the blob can be localised for the given credibility level $1-\alpha$. For example, in figure \ref{fig:significant_blobs} we see that two blobs on the right correspond to the same blue circle. This means that we cannot separate them with credibility above $1 - \alpha$. We are confident that there is at least one blob in the true image $\f$ that corresponds to the blue circle, but we cannot say with credibility that there are two. Hence, these circles can be used as a tool to visualise localisation uncertainty in the blob detection.
}

\subsubsection{Matching procedure}\label{sssec:matching}

We have not yet described the details of the matching procedure mentioned in the description of ULoG. There are of course many possibilities to define a criterion for blobs "matching". For our purposes, the following procedure proved successful:
\begin{enumerate}
\item First, the detected blobs $\mathcal B$ in the MAP estimate are \removed{cleaned}\added{post-processed, as described in section \ref{subsec:laplacian_of_gaussians}. Which blobs remain in $\mathcal B$ depends on the strength threshold $l^\mathrm{thresh}$ and the overlap threshold $o_1^\mathrm{thresh}$.} \removed{Let $\ssrmap$ be the scale-space representation of $\fmapvec$. All blobs for which the value of $\normlap \ssrmap$ is above a certain threshold $l^\mathrm{tresh}$ are removed (since the strength of a feature is indicated by a strongly \emph{negative} value of $\normlap \ssrmap$). In our experiments, we used a relative treshold of the form $l^\mathrm{tresh} = r^\mathrm{tresh} \cdot (\min \normlap \ssrmap)$, where $r^\mathrm{tresh} \in (0, 1)$.
\item The relative overlap of the detected blobs is computed. Hereby, a blob $\bm b = (i, j, q)$ is identified with a circle of radius $2 \sqrt{t_q}$, centered at the pixel $(i, j)$. If the relative overlap of two blobs surpasses a treshold $o^\mathrm{tresh}_1$, the blob with the larger value of $\normlap \ssrmap$ is removed.}
\item The scale-space minimisers $\mathcal C$ of $\bm \normlap \blanket$ are cleaned in the same way, the only difference being that $\bm \normlap \ssrmap$ is replaced with $\bm \normlap \blanket$.
\item Now, for every $\bm b \in \mathcal B$ we check whether there is an overlap with any $\bm c \in \mathcal C$. Here, overlap is defined in the same way as in step $(ii)$. If there exists a $\bm c \in \mathcal C$ such that its relative overlap with $\bm b$ is above a treshold $o^\mathrm{tresh}_2$, then blobs $\bm b$ and $\bm c$ are a "match".
\end{enumerate}

The ideal configuration for the tunable parameters in this procedure will typically depend on the application. In our numerical experiments, we obtained satisfying results using \removed{the values $r^\mathrm{tresh} = 0.05$}\added{$l^\mathrm{thresh} = r^\mathrm{thresh}\cdot \abs{\min \normlap \ssrmap}$ with $r^\mathrm{thresh} = 0.02$} and $o^\mathrm{tresh}_1 = o^\mathrm{tresh}_2 = 0.5$.

\added{
\subsubsection{Visualisation}\label{sssec:visualise}
}

There are multiple ways in which we could visualise the results of the outlined algorithm. In figure \ref{fig:significant_blobs}, the following procedure was applied:
\begin{itemize}
\item A scale-space blob $\bm b = (i, j, q)$ is visualised as a circle with center $(i, j)$ and radius $\sqrt{2 t_q}$ (see section \ref{subsec:laplacian_of_gaussians}).
\item If a blob $\bm b \in \mathcal B$ \added{does not match a blob in $\mathcal C$}\removed{is "not significant"}, it is visualised as a red circle.
\item If a blob $\bm b \in \mathcal B$ \added{matches a blob in $\mathcal C$}\removed{is "significant", i.e. if there is a matching blob $\bm c \in \mathcal C$}, it is visualised as a green circle. Furthermore, the match $\bm c$ is visualised as a dashed, blue circle.
\end{itemize}

The dashed blue circles have to be interpreted in the following way: If there is a second structure inside the circle, then we cannot distinguish it with confidence from the first blob. Hence, the radius of the circle indicates how well the blob can be localised for the given credibility level $1-\alpha$. For example, in figure \ref{fig:significant_blobs} we see that two blobs on the right correspond to the same blue circle. This means that we cannot separate them with credibility above \removed{$1 - \alpha$}\added{95\%}. We are confident that there is at least one blob in the true image $\f$ that corresponds to the blue circle, but we cannot say with credibility that there are two. Hence, these circles can be used as a tool to visualise localisation uncertainty in the blob detection.

\section{Application to M54}\label{sec:m54}

\begin{figure*}
\resizebox{\hsize}{!}{\includegraphics{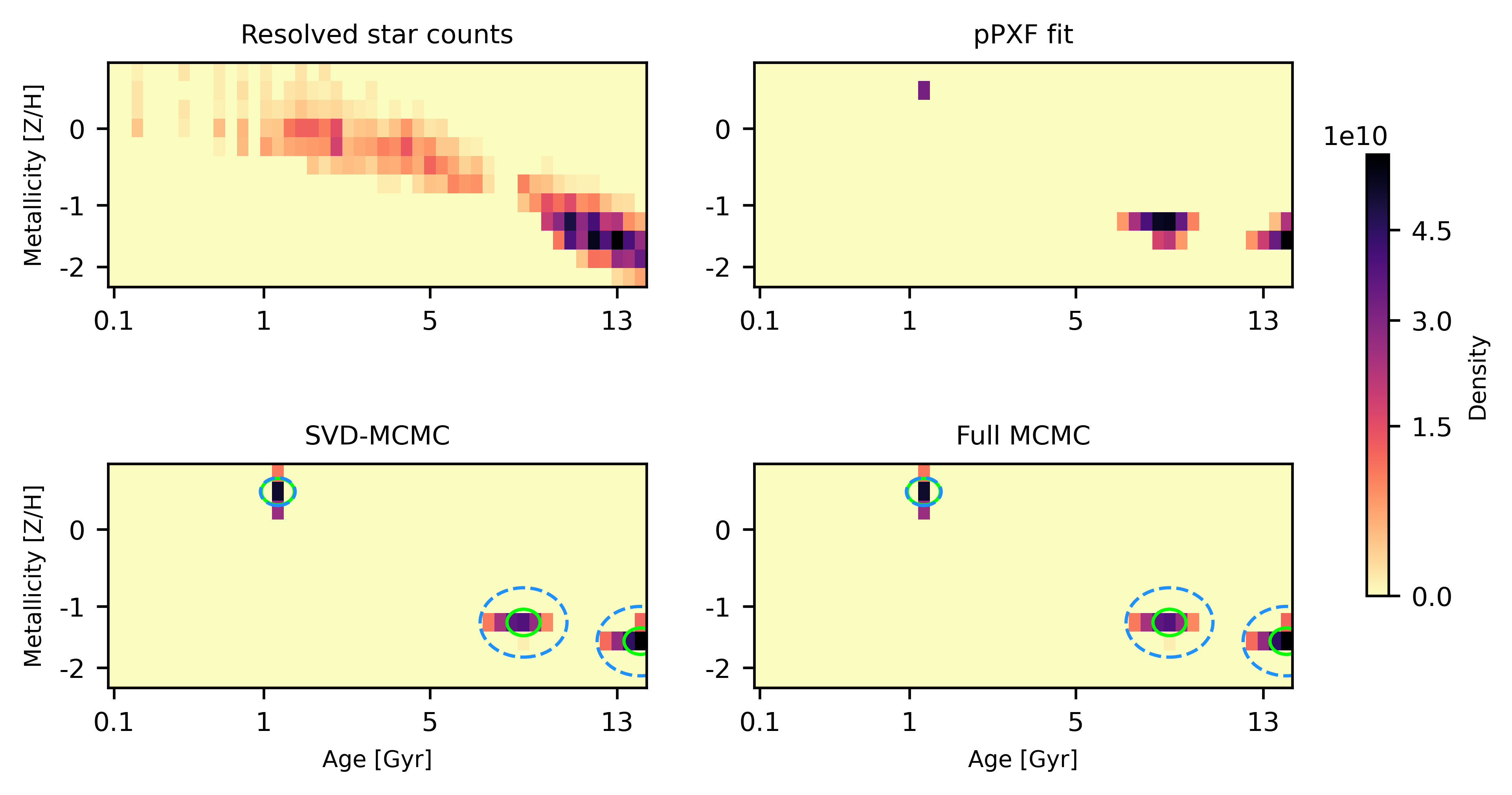}}
      \caption{Age-metallicity distributions for the M54 star cluster.
			Top left: the ground truth number density from individual measurements of resolved stars in M54, from \citet{AlfaroCuello19}. Top right: a MAP revovery using PPXF full spectrum fitting of an integrated-light M54 spectrum, from \citetalias{BoeAlfNeuMarLea20e}.
			Bottom row: Results of ULoG on M54 data using low-regularisation. Both panels show the same MAP estimate of the recovered posterior probability density. The coloured circles show the location of significant blobs with 95\%-credibility, with colours as in Figure \ref{fig:significant_blobs}. The left shows results using SVD-MCMC, the right shows Full MCMC. In both cases, ULoG identifies 3 distinct blobs with 95\%-credibility.}
      \label{fig:m54_blobs_lowreg}
\end{figure*}

M54 is the nuclear star cluster of the Sagittarius dwarf galaxy and hosts a mixture of stellar populations.
\citet{AlfaroCuello19} studied these populations using resolved data of individual stars in the cluster.
They combined photometry from the Hubble Space Telescope with spectroscopic measurements from the Multi-Unit Spectroscopic Explorer (MUSE) to measure ages and metallicities of 6600 M54 stars, whose distribution is shown in the top-left panel of Figure \ref{fig:m54_blobs_lowreg}.
Two distinct clumps are clearly identifiable: an old metal-poor clump and young metal-rich clump, both of which display a trend of increasing metallicity for younger stars.
The authors hypothesised that the old metal-poor population is formed from the accretion and merger of old and metal poor globular clusters, while the young population is some combination of stars formed in-situ in M54 and interlopers from the Sagittarius dwarf.
We test our blob detection techniques on M54, with these resolved measurements providing a \emph{ground truth} distribution to compare with.

\citet{BoeAlfNeuMarLea20e} (hereafter B20) used the M54 MUSE data to compare age-metallicity recoveries from resolved measurements against those from integrated spectra. They created an integrated M54 spectrum (SNR = 90) by co-adding the spectra of individual M54 stars and analysed this using pPXF. The resulting recovery is shown in the upper right panel of Figure \ref{fig:m54_blobs_lowreg}. It matches the ground-truth distribution in coarse statistics, with mean age and metallicities from the two distributions separated by just 0.3 Gyr and 0.2 dex $[\mathrm{Fe}/\mathrm{H}]$ respectively. The integrated recovery also displays distinct populations, well separated in age and metallicity. We see a young metal-rich population (single bright pixel at the top), and old metal-poor contributions also.

One further difference between the \citetalias{BoeAlfNeuMarLea20e} result and the ground truth distribution is that the integrated recovery splits the old population into two clumps. The authors quantified the uncertainty in their pPXF recovery by Monte-Carlo resampling the data 100 times, randomly perturbed by the residual from the baseline pPXF fit. They calculated 100 MAP estimates from the resampled data, interpreting these as posterior samples. The pixel-wise 84\% credible intervals of these samples enveloped the two clumps in the old population, suggesting that the split may not be significant.
Our uncertainty-aware blob detection techniques are well placed to address this question in a principled way: our MCMC samples are directly drawn from the posterior, and ULoG is a well-motivated way to do blob detection using these samples. We therefore apply these techniques to the integrated M54 spectrum from \citetalias{BoeAlfNeuMarLea20e} to assess the significance of the recovered blobs.

\subsection{Notes on the previous works}\label{ssec:m54_caveats}

In the \citet{AlfaroCuello19} resolved study, the authors dissect the metal-rich clump into two distinct populations. Whilst this extra split is supported by spatial and kinematic information \citep{AlfaroCuello20}, based purely on the age-metallicity distribution seen in the upper left panel of Figure \ref{fig:m54_blobs_lowreg} this split is not visible. \removed{We therefore do not consider this extra spilt in our results.}

We noticed a small systematic effect in the \citetalias{BoeAlfNeuMarLea20e} integrated-light study. In their recovery, a non-zero weight was assigned to the youngest SSP (0.03 Gyr) at an intermediate metallicity, with no counterpart in the resolved distribution. This very young contibution was not visible in the \citetalias{BoeAlfNeuMarLea20e} paper since they use a linear axis for age, squashing very young SSPs to an inperceptably narrow range. Our visualisation - with a image pixel per SSP - highlighted this spurious contibution much more significantly. We believe this is a systematic effect. The spectra of the very youngest SSPs, hosting the hottest stars, are relatively featureless i.e. showing relatively few and weak absorption features. A small contribution of very young stars may therefore be used to offset residuals from incorrect flux calibration and/or treatment for dust absorption. In order to focus our experiments on blobs with counterparts in the ground truth, we therefore exclude SSPs below 0.1 Gyr in our recoveries. After this cut, no weight is allocated to the youngest available SSP (i.e. 0.1 Gyr), supporting the hypothesis that this very young contribution was a systematic effect.

\subsection{Setup}\label{ssec:m54_setup}

The inputs for our recoveries mirror those used in \citetalias{BoeAlfNeuMarLea20e}. We use the same library of SSP templates i.e. E-MILES library \citet{Vazdekis16}, with an bimodal IMF of slope 1.3, BaSTI isochrones \citep{Pietrinferni04} using all available metallicities, but only retaining ages above 0.1 Gyr (see Section \ref{ssec:m54_caveats}). We mask the wavelengths ranges $[5850,5950], [6858.7-6964.9]$ and $[7562.3-7695.6]$ (Angstrom) to avoid large sky-residuals. Since our analysis routines are specialised to the case of the purely linear part of spectral modelling, we first perform a fit with pPXF and fix the non-linear parts to the pPXF best-fit values. We find best fitting kinematic parameters of $[V=146. \mathrm{km/s}, \sigma=3. \mathrm{km/s}, h_3=0.088, h_4=0.187]$ and the multiplicative polynomial of order 21 used by pPXF for continuum correction. We pre-process the SSP templates to account for these effects, convolving them with the LOSVD and multiplying them by the polynomial. Our Bayesian model was set up using a truncated Gaussian prior \eqref{eq:truncated_gaussian_prior} with an Ornstein-Uhlenbeck covariance \eqref{eq:ornstein_uhlenbeck}.
The M54 integrated spectrum and noise estimates were provided by the \citetalias{BoeAlfNeuMarLea20e} authors. We try two choices for the regularisation parameter $\beta$: a \emph{low regularisation} choice which gives a MAP reconstruction similar to the \citetalias{BoeAlfNeuMarLea20e} result ($\beta=1 \cdot 10^3$), and a \emph{high regularisation} option for comparison ($\beta=\removed{1}\added{5} \cdot 10^5$).

We make the following specific choices for ULoG. To compute filtered credible intervals, we  Full MCMC and SVD-MCMC with $q=15$ (chosen as described in section \ref{sssec:svd_choose_q}). In both cases we follow the procedure laid out in section \ref{sssec:mcmc_implementation} i.e. taking 10000 samples for $\f$ and checking that convergence diagnostics are satisfied. To compute filtered credible intervals we consider 10 scales $t_k = 1 + (k-1) \cdot 1.5$. The computation times \added{(using Intel MKL on 8 3.6-GHz Intel Core i7-7700 CPUs and 64 GB RAM)} were as follows: for low/high regularisation, SVD-MCMC required 1785/998 seconds, while Full MCMC required 12908/5295 seconds. This implies a speedup factor of roughly 5 to 10 for SVD-MCMC over the full model, with greater speedup for less regularisation.

\subsection{Results}\label{ssec:m54_results}

\begin{figure*}
\resizebox{\hsize}{!}{\includegraphics{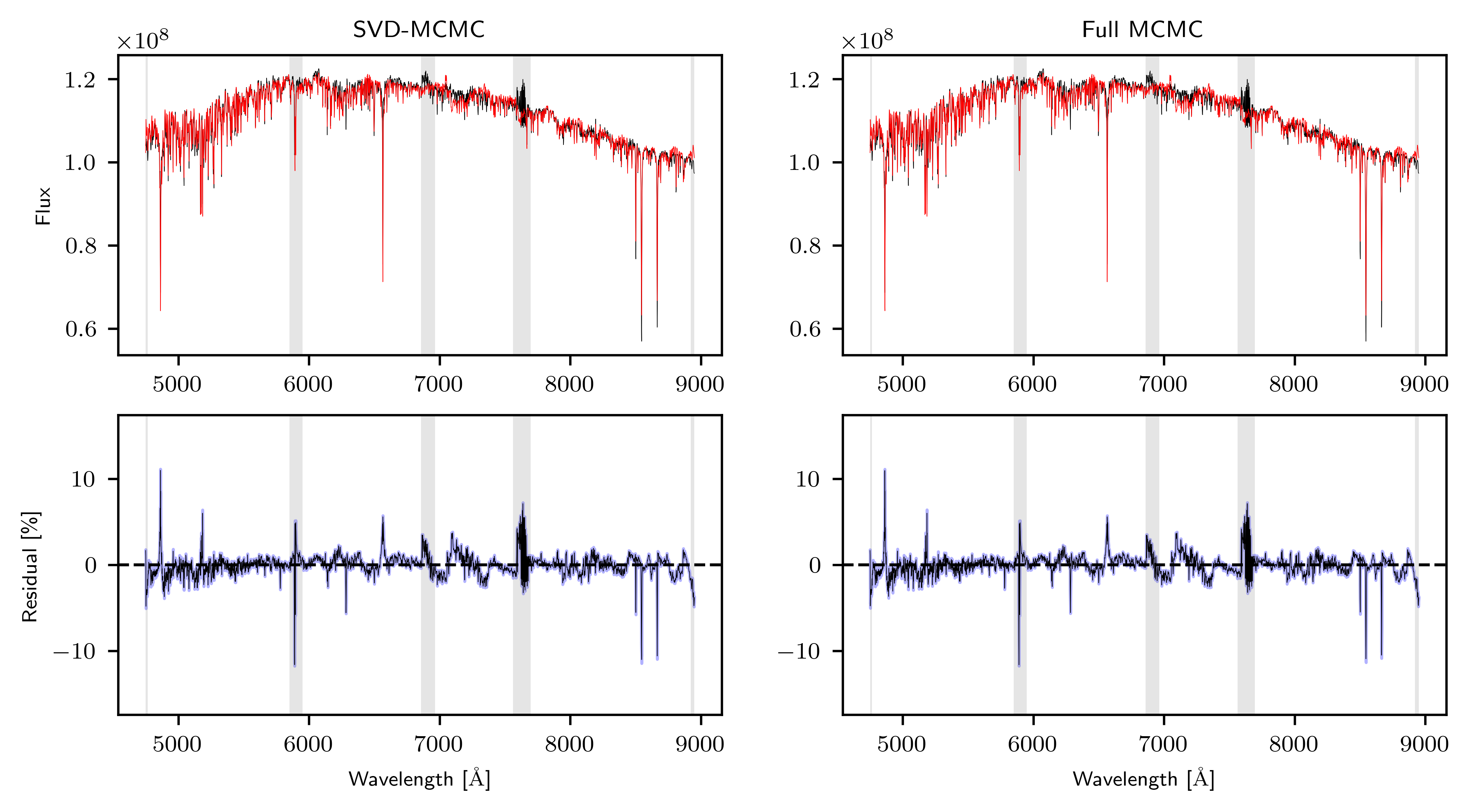}}
\caption{Plots of the integrated M54 spectrum $\obs$ and the data fit $\obs^\text{pred}$ (in red), both for SVD-MCMC (left) and Full MCMC (right). The data fit is related to the respective MCMC posterior mean $\f^\text{mean}$ through $\bm y^\text{pred} = \G \f^\text{mean}$. The plots in the second row show the relative residuals $r^\text{rel}_i = (y_i - y^\text{pred}_i) / y_i$, where $\bm y$ is the measurement data. The corresponding band (in blue) shows the simultaneous posterior predictive 95\%-credible intervals. Grey vertical bands show wavelength regions which have been masked to avoid sky-residuals.}
\label{fig:posterior_predictive}
\end{figure*}

Fits to the data for the low-regularisation case are shown in Figure \ref{fig:posterior_predictive}. The residuals are smaller than \removed{0.1\%} \added{10\%}, showing that the data is well-fit. The fact that residuals are correlated between neighbouring wavelengths suggests inaccuracies at the \removed{0.1\%} \added{10\%} level in the assumed, uncorrelated noise model. The similarity between the SVD and Full-MCMC data-fits gives us confidence that the SVD truncation has not biased the results. The data fits for the high-regularisation case look very similar.

\begin{figure}
\centering
\includegraphics[width=\hsize]{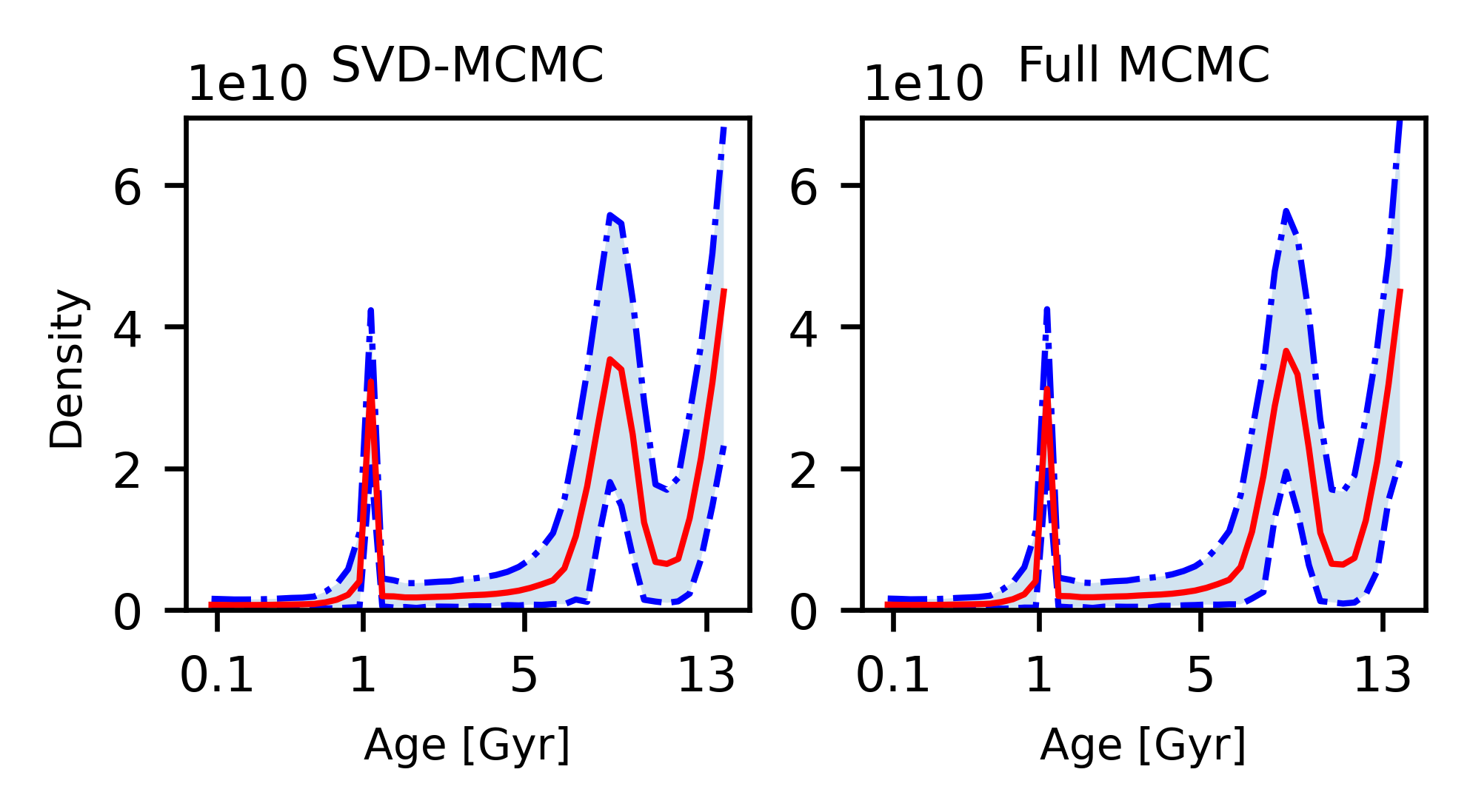}%
\caption{Marginal posterior for the age distribution\added{ (see also appendix \ref{subsec:age_marginal})}, for the low-regularisation case. We show 95\%-credible intervals (shaded area) and marginal posterior mean (solid red line). Left: Based on SVD-MCMC. Right: Based on Full MCMC.}
\label{fig:m54_marginal_lowreg}
\end{figure}

The recoveries for the low-regularisation case are shown in the bottom panels of Figure \ref{fig:m54_blobs_lowreg}.
Results for both SVD-MCMC and Full MCMC again agree well.
In both cases, ULoG detects three distinct blobs: one young and metal rich, and two old and metal poor.
Importantly, we infer that these two old and metal poor blobs are not likely to be part of a single, combined blob.
Justification for this can be seen clearly in Figure \ref{fig:m54_marginal_lowreg} which shows marginalised constraints on the star formation histories: for old ages >5 Gyr, the 95\% credible band are inconsistent with a single blob.
For this level of regularisation, our computations clearly separate the old contribution into two metal-poor blobs whereas \citetalias{BoeAlfNeuMarLea20e} does not.
This conclusion strongly depends on the regularisation used, however.

\begin{figure*}
\resizebox{\hsize}{!}{\includegraphics{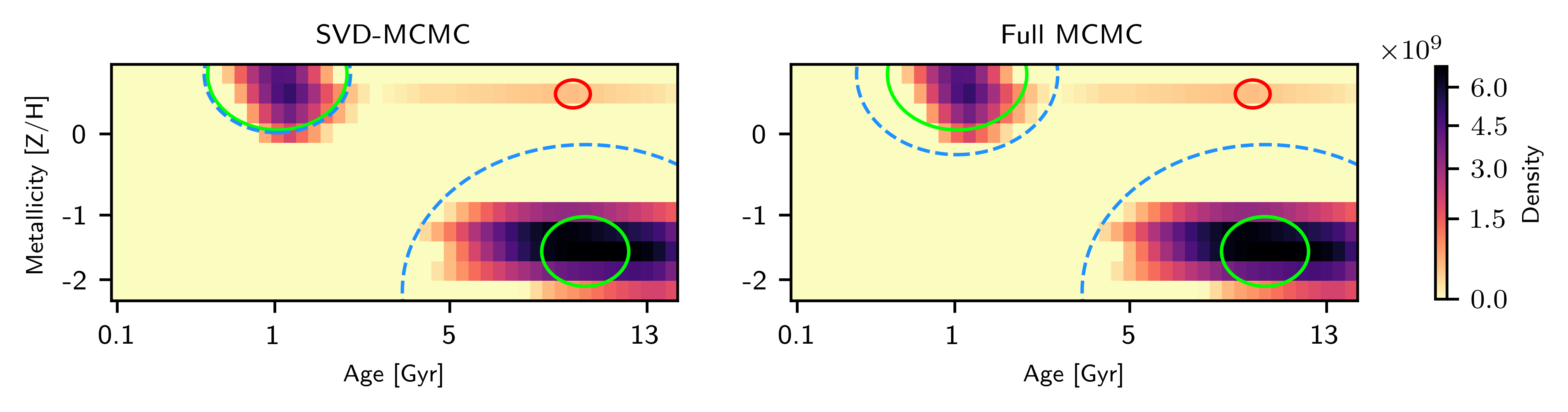}}
      \caption{Results of ULoG on M54 data with higher regularisation.}
 \label{fig:m54_blobs_highreg}
\end{figure*}

\begin{figure}
\centering
\includegraphics[width=\hsize]{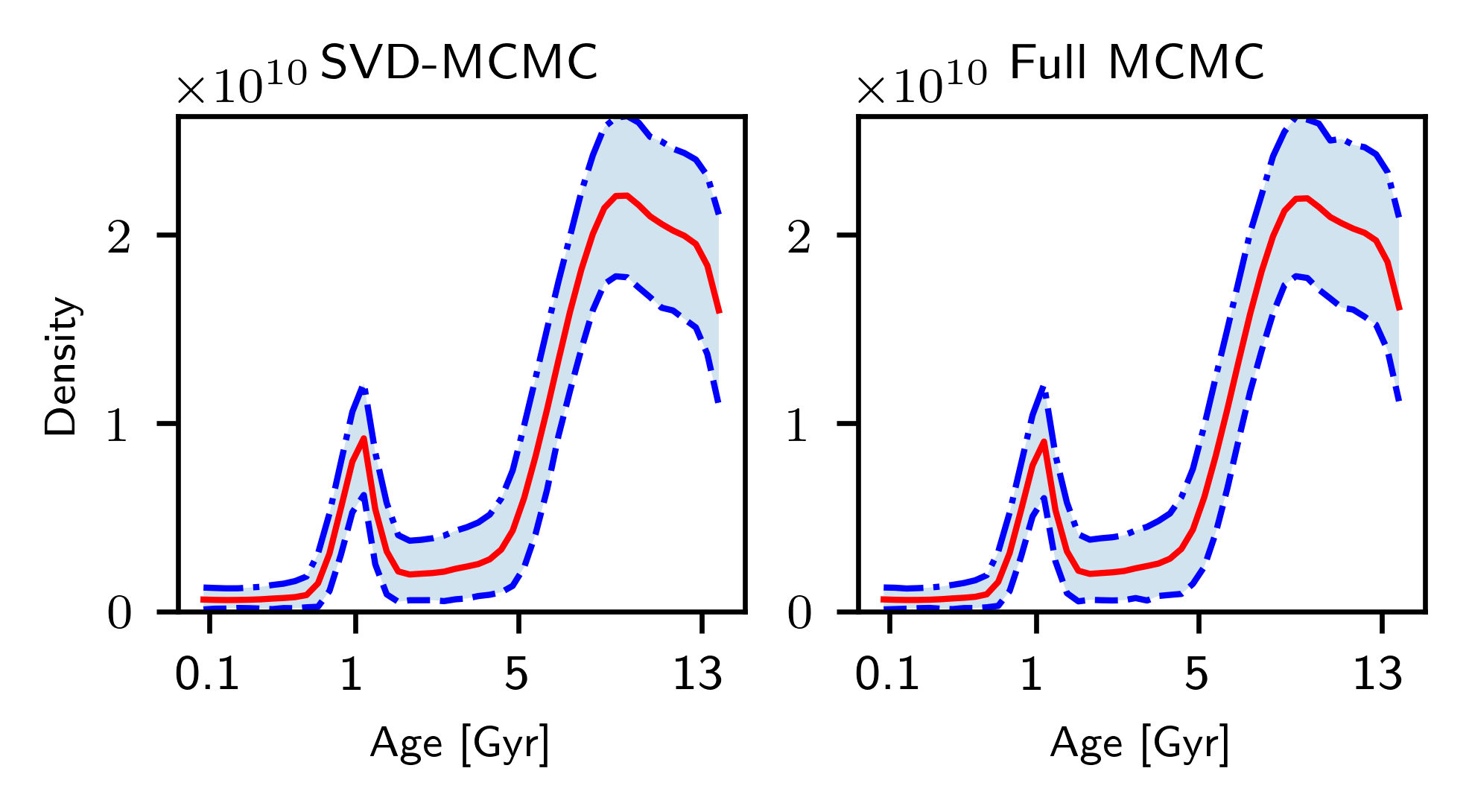}%
\caption{As for Figure \ref{fig:m54_marginal_lowreg} but with higher regularisation.}
\label{fig:m54_marginal_highref}
\end{figure}

Figures \ref{fig:m54_blobs_highreg} and \ref{fig:m54_marginal_highref} show the inferred distributions using the high-regularisation option.
Though the young and metal rich blob is still detected as significantly distinct, the two old and metal poor blobs have now merged together.
Additionally, an insignificant blob (red circle) is now seen at an old and metal-rich population. The origin of this third blob is unclear, however it is not present when we extend the SSP grid to include younger ages. This is likely to be some systematic effect, however it is small and ultimately deemed statistically insignificant by ULoG.
This detection of only two significant blobs more accurately matches the ground truth (top left panel of Figure \ref{fig:m54_blobs_lowreg}) which displays two rather than three distinct blobs.
As with the low-regularisation case, in both Figures \ref{fig:m54_blobs_highreg} and \ref{fig:m54_marginal_highref} we see that the results of both SVD-MCMC and Full MCMC agree well.
Thus, while our experiments have hopefully demonstrated the utility of ULoG (sped up with an SVD-MCMC backend) the dependence of blob detection on regularisation parameter remains an open question, which is discussed further in the next section.

\section{Discussion}\label{sec:discussion}

The ideas explained in section \ref{sec:uncertainty_aware_blob_detection} relate concepts from different fields such as mathematics, statistics and image processing. Since ULoG uses scale-space representations to perform statistical inference, it can be viewed as an example of a statistical scale-space method \citep[see e.g.][]{HolPas17}. To our knowledge, the uncertainty-aware scale-space representation based on "blankets" (see section \ref{subsec:ulog}) is new to the literature. Nevertheless, similar ideas have been used in variational imaging for a long time, although not in the context of uncertainty quantification, and are known as "taut string" or "tube methods" \citep[see][]{MamGee97, DavKov01, HinHinKunOehSch03, SchGraGroHalLen09}. Clarifying all of these connections is a promising starting point for theoretical work, since the resulting insights could be used to address various inference problems in astronomical imaging.

Our implementation of ULoG is based on SVD-MCMC (see section \ref{sec:svd_mcmc}). In this case, the overall computational cost of the method is dominated by the generation of posterior samples. Given these samples, the additional cost of computing filtered credible intervals for a range of scales is negligible. Having access to posterior samples, one might also consider alternative ways to perform uncertainty-aware blob detection. One could try to circumvent the computation of filtered credible intervals altogether and instead apply the LoG method (or any other known blob detection algorithm) directly to each sample. However, such a direct approach has to tackle a number of challenges: First, there is no obvious way how to combine the results of the blob detection from each individual sample into meaningful statistics. Different samples might have blobs of widely differing size at non-matching locations. Furthermore, individual posterior samples tend to be quite noisy. This is problematic, since a blob detection method applied to a noisy sample will detect a large number of "spurious" blobs, making it very challenging to extract meaningful information from its result.

However, ULoG has the key advantage that it only requires computation of filtered credible intervals. This means that it does not explicitly need access to posterior samples and hence can be combined with any method of estimating filtered credible intervals. For example, in the context of linear imaging problems \cite{Per17} proposed to estimate \removed{the }posterior credible region\added{s} using concentration-inequalities, circumventing the need for sampling. Approaches based on this idea have been shown to be very scalable \citep[see e.g.][]{CaiPerMcE18a}. Using concentration-inequalities to directly estimate filtered credible intervals could make ULoG applicable to very large imaging problems. We will investigate this idea in future work.

\subsection{SVD-MCMC}\label{subsec:svd_mcmc_discussion}

We have seen in section \ref{sec:m54} that the choice of the regularisation parameter $\beta$ in equation \eqref{eq:prior} affects the inferences. There are various strategies to address the choice of $\beta$ in the Bayesian framework, which would be straightforward to combine with SVD-MCMC. In a hierarchical approach \citep[see e.g.][]{MolKatMat99, PerBioFig15}, we could assign a prior on $\beta$. Alternatively, we could estimate $\beta$ from the data using empirical Bayesian methods \citep[e.g. as in][]{PerAntLec15, VidBorPerDur20}. We could also try to eliminate the presence of the regularisation parameter altogether, and use a simulation-based prior \citep[for an example in the context of weak gravitational lensing, see ][]{Rem22}.

While in this paper we implemented SVD-MCMC with HMC, one could consider different MCMC methods. For example, SVD-MCMC could be combined with specialised samplers for nonsmooth probability densities, such as proximal MCMC \citep[][]{DurMouPer18}, to allow e.g. the use of $\ell^1$-Laplace priors, which is a common choice in imaging \citep[see e.g.][]{CasChaNov15}.

In our numerical tests, we observed that the HMC sampler ran into numerical problems when the regularising prior was assigned to $\f$ in mass-weights instead of light-weights. Specifically, a significant number of divergent transitions occurred \citep{Betancourt16}, and $\hat{R}$ statistics were much greater than 1, signalling unconverged chains. This is likely due to the fact that the observation operator $\mathcal G$ is much more ill-conditioned with respect to mass-weighted distribution functions. This problem could be resolved by replacing HMC with a more robust sampler, such as e.g. Randomize-Then-Optimize \citep[][]{BarSolHaaLai14}, though it is unclear how efficient this specific option would be for such a high-dimensional problem.

\subsection{Applications beyond the current problem}\label{subsec:further_applications}

In this work we have presented ULoG and SVD-MCMC on an idealised problem. Several other effects may be important for spectral modelling, including convolution with the LOSVD, treatment for reddening, and accounting for gas emission. For the M54 experiment presented here, we have pre-processed the data to remove these effects before proceeding with the linear stellar population model. We note that in the future, these other effects could be incorporated into a more complete model which is still fully compatible with SVD-MCMC and ULoG.

Our method for uncertainty-aware blob detection is very general and may find applications beyond spectral modelling. One specific example which we intend to explore in the future is the task of blob detection in orbit distributions inferred from orbit-superposition modelling of galaxies \citep[e.g.][]{Zhu22}.

In our application to M54, we have seen that SVD-MCMC can lead to a reduction of computation time up to a factor 7 compared to Full MCMC. For larger problems, this factor might even increase as the full potential of the truncated SVD comes truly into play in higher dimensions. Potentially, this could make uncertainty quantification feasible not only for single-spectrum fitting, but even for full IFU datacube modelling \citep[see e.g.][]{HinHubJetSooVen23}.

\section{Conclusions}\label{sec:conclusions}

In this work we have adressed the question of blob detection in images with uncertainties. We have proposed a novel, uncertainty-aware variant of the classic Laplacian-of-Gaussians method for blob detection (ULoG). ULoG detects significant blobs considering various scales, and requires as input uncertainty estimates for the image in the form of filtered credible intervals. Furthermore, we describe an efficient computational method for calculating filtered credible intervals which combines truncated SVD and MCMC sampling (SVD-MCMC).

These methods have been developed in the context of stellar population modelling of integrated-light spectra. We illustrated the methods first on mock data, and then applied it to real data of the nuclear star cluster M54. Results using SVD-MCMC and Full MCMC match well. The observed speed-up using SVD-MCMC is on average a factor 7, with sampling lasting 15-30 minutes as opposed to 2-4 hours. The number of significant blobs detected ranged between 2-3, depending on the amount of regularisation used. This work focussed on uncertainty-aware blob detection at fixed regularisation, but in the future our methods can be combined with Bayesian techniques for selecting or marginalising over the regularisation parameter.

\begin{acknowledgements}

\added{The authors thank the referee for a constructive and helpful report.}
PJ and GvdV acknowledge funding from the European Research Council (ERC) under the European Union's Horizon 2020 research and innovation programme under grant agreement No 724857 (Consolidator Grant ArcheoDyn). PJ and GvdV also acknowledge \removed{supoort}\added{support} by the Austrian Science Fund (FWF): F6811-N36. FP and OS were funded by the Austrian Science Fund (FWF): F6807-N36.
The financial support by the Austrian Federal Ministry for Digital and Economic
Affairs, the National Foundation for Research, Technology and Development and the Christian Doppler
Research Association is gratefully acknowledged.
AB was funded by the Deutsche Forschungsgemeinschaft (DFG, German Research Foundation) -- Project-ID 138713538 -- SFB 881 (``The Milky Way System'', subproject B08) and acknowledges support from grant ProID2021010080 in the framework of Proyectos de I+D por organismos de investigaci\'{o}n y empresas en las \'{a}reas prioritarias de la estrategia de especializaci\'{o}n inteligente de Canarias (RIS-3), FEDER
Canarias 2014-2020.

\end{acknowledgements}

\bibliographystyle{aa}
\bibliography{ParJetBoeAlfSchVen23}

\begin{appendix}

\added{
\section{Simultaneous vs. pixel-wise credible intervals}\label{app:simultaneous_vs_pixelwise}

In section \ref{subsec:credible_intervals} we defined credible intervals as \emph{simultaneous} intervals. These need to be distinguished from \emph{pixel-wise} credible intervals that are also often used for uncertainty visualisation (for scalar parameters, these notions coincide). Given a pixel $(i, j)$, we call an interval $[\ell_{ij}, u_{ij}] \subset \R$ a \emph{pixel-wise posterior $(1-\alpha)$-credible interval} if it is a $(1-\alpha)$-credible interval with respect to the \emph{marginalised} posterior distribution $p(f_{ij} | \obs)$. That is, if there holds
\begin{equation*}
P(f_{ij} \in [\ell_{ij}, u_{ij}] | \obs) \geq 1 - \alpha.
\end{equation*}
If we assemble all lower bounds $(\ell_{ij})$ and upper bounds $(u_{ij})$ into images $\bm \ell, \bm u \in \R^{T \times Z}$, these images define then a multi-dimensional interval $[\bm \ell, \bm u] \subset \R^{T \times Z}$ via equation \eqref{eq:multidimensional_interval}. However, this interval will in general not satisfy equation \eqref{eq:simultaneous_credible_intervals}, as pointed out e. g. by \cite{Gan09}. For our purposes, simultaneous credible intervals are more suitable, since by definition they satisfy equation \eqref{eq:simultaneous_credible_intervals}, thus allowing the intuitive interpretation that the signal of interest lies "between" the lower bound $\flow$ and the upper bound $\fupp$ with posterior probability greater or equal $1 - \alpha$. 

The difference between simultaneous and pixel-wise credible intervals can also be stated by referring to the concept of  \emph{posterior credible regions}.  We call a set $\region \subset \R^{T \times Z}$ a posterior $(1-\alpha)$-credible region for $\f$ if
\begin{displaymath}
p(\f \in \region | \obs) \geq 1 - \alpha.
\end{displaymath}
By equation \eqref{eq:simultaneous_credible_intervals}, a simultaneous $(1-\alpha)$-credible interval is always a $(1-\alpha)$-credible region, while this might fail for a pixel-wise $(1-\alpha)$-credible interval.

We also remark that the terminology in this area is not completely standardised. Credible regions and credible intervals are sometimes called Bayesian confidence regions and Bayesian confidence intervals, respectively, while the multi-dimensional simultaneous credible intervals mentioned above have been referred to as Bayesian confidence bands or credible bands in the area of time series analysis \citep[see][]{Bre78}.

\subsection{Credible intervals for the age distribution}\label{subsec:age_marginal}

The credible intervals for the age distribution plotted in figures \ref{fig:m54_marginal_lowreg} and \ref{fig:m54_marginal_highref} were computed as follows: For each MCMC sample $\f^s$, we sum over the metallicity,\begin{align*}
a_j^s = \sum_{i=1}^Z f_{ij}^s, \qquad s=1, \ldots, S.
\end{align*}
Then, we estimated simultaneous $95\%$-credible intervals for $\bm a \in \R^T$ using the method outlined in section \ref{subsec:fcis_from_samples} (that is, setting $\bm x^s = \bm a^s$).
}

\section{Choice of $q$}\label{sssec:svd_choose_q}

\begin{figure}
   \centering
   \includegraphics[width=\hsize]{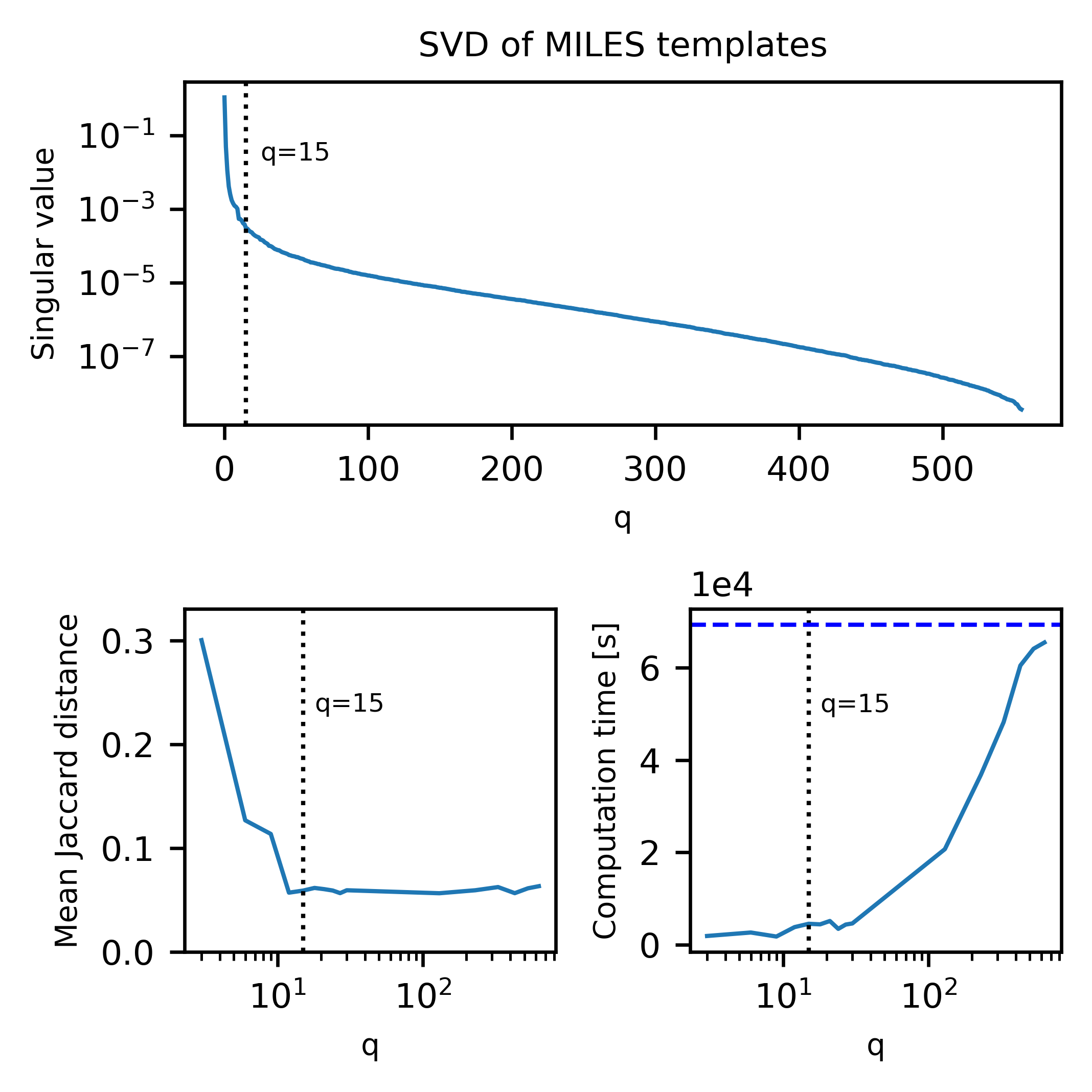}
      \caption{The influence of the latent dimension $q$ on SVD-MCMC. The top panel shows singular values of the matrix of MILES SSPs. The lower left panel shows the error between filtered credible intervals computed with SVD-MCMC and Full MCMC. The error is measured in the mean Jaccard distance (see equation \eqref{eq:mean_jaccard}). Beyond $q=15$ the error stabilises, signalling converged results. The lower right panel shows the corresponding computation time for SVD-MCMC, where the dashed horizontal line marks the computation time for Full MCMC. In all panels, the vertical dotted line is at $q=15$, the value adopted for SVD-MCMC inferences in this work.}
      \label{fig:svd}
\end{figure}

How can we choose an appropriate size of the latent dimension $q$? One possibility is to look at the rate of decay of singular values. These are shown for our problem in the top panel of Figure \ref{fig:svd}. Singular values decay more slowly beyond $q=15$, hence the accuracy of the truncated SVD approximation grows more slowly beyond this value. Choosing a value of $q$ around this inflection point provides a good balance between accuracy and possible speed-up from dimensionality reduction. Note however that this choice is not \emph{guaranteed} to be sufficient in all circumstances e.g. for very high signal-to-noise data it is conceivable that a larger $q$ may be required. We therefore recommend to check convergence of results against the choice $q$.

To assess convergence we will run SVD-MCMC over a range of $q$ and quantify changes in the results. The metric we will use to compare two sets of MCMC samples is the \emph{mean Jaccard distance} between filtered credible intervals computed from the samples. For two intervals $[u^1, v^1], [u^2, v^2] \subset \R$, we define their \emph{Jaccard distance} as
\begin{displaymath}
\quad  D_J([u^1, v^1], [u^2, v^2]) = 1 - \frac{\abs{[u^1, v^1] \cap [u^2, v^2]}}{\abs{[u^1, v^1] \cup [u^2, v^2] }}.
\end{displaymath}
The value of the Jaccard distance is always between $0$ and $1$, where $0$ corresponds to identity and $1$ corresponds to disjointness (up to sets of measure zero).
To measure the distance between filtered credible intervals $[\bm{u}^1, \bm{v}^1], [\bm{u}^2, \bm{v}^2] \subset \R^{T \times Z}$ in image space, we take the mean Jaccard distance over all pixels:
\begin{equation}
\overline D_J([\bm{u}^1, \bm{v}^1], [\bm{u}^2, \bm{v}^2]) := \frac{1}{T \cdot Z} \sum_{i=1}^T \sum_{j=1}^Z D_J([u^1_{ij}, v^1_{ij}], [u^2_{ij}, v^2_{ij}]). \label{eq:mean_jaccard}
\end{equation}

The lower left panel of figure \ref{fig:svd} shows the results of our convergence test on mock data. It shows the mean Jaccard distance between filtered credible intervals computed using Full MCMC and SVD-MCMC with 5000 warmup and 5000 sampling steps as a function of $q$. The filtered credible intervals were computed for scale $t=5$ and credibility parameter $\alpha=0.05$, but we observed that the choice of these parameters did not significantly affect the resulting plot. Beyond $q=15$ the distance does not decrease, signalling that filtered credible intervals do not systematically change and our results are converged. The minimum distance does not reach zero, but instead flattens around $0.05$ due to the inherent stochasticity of samples (i.e. two independent sets of samples from the Full MCMC model would also show this level of difference). The bottom right panel of figure \ref{fig:svd} shows the computation time required for sampling the SVD-MCMC models as a function of $q$. At $q=15$, we still benefit from a factor 15 speedup compared to sampling the full model.

For general usage of SVD-MCMC (where it may be infeasible/undesirable to compare against Full MCMC), we suggest the following strategy for picking $q$. Pick some initial $q_0$ smaller than the inflection point of the singular values and a step-size $\Delta q\approx 5$. Perform SVD-MCMC inferences at $q=q_0+i\Delta q$, first for $i=0$ and $1$, and compute the Jaccard distance between these two sets of results. If $q_0$ is small enough, there will likely be a significant decrease in Jaccard distance between $i=0$ and $1$. Finally, increment $i$, computing the Jaccard distance between the cases $i$ and $i-1$, till the distance stabilises to a constant value.

\end{appendix}

\end{document}